\documentclass[journal]{IEEEtran}

\IEEEoverridecommandlockouts      

\usepackage{amsmath}
\usepackage{epsfig,tabularx,amsfonts,amssymb}
\usepackage{amsfonts}
\usepackage{graphics} 
\usepackage{epic,eepic,epsfig}
\usepackage{mathrsfs}
\usepackage{subfigure}
\usepackage{calc,ifthen}
\usepackage{color}
\usepackage{bm}
\usepackage{float}
\usepackage{pstricks}
\usepackage[normalem]{ulem}
\usepackage{hyperref}
\usepackage[capitalize, nameinlink]{cleveref}
\newcommand{\stkout}[1]{\ifmmode\text{\sout{\ensuremath{#1}}}\else\sout{#1}\fi}

\usepackage{enumitem}
\usepackage{comment}

\newtheorem{problem}{\upshape OCP}

\newtheorem{remark}{Remark}
\newtheorem{assumption}{Assumption}

\def\lungo #1{\mathord{\buildrel{\lower3pt\hbox{$\scriptscriptstyle\frown$}}
\over #1 } }

\def\salt{\vskip 2.4 true mm}

\def\1D#1#2{{{\partial}\over{\partial #2}}#1}

\def\1d#1#2{{{d}\over{d #2}}#1}

\newcommand{{\R}}{{\mathbb{R}}}
\newcommand{{\C}}{{\mathbb{C}}}

\usepackage{fancyhdr}

\title{A Convex Optimal Control Framework for Autonomous Vehicle Intersection Crossing}

\author{Xiao Pan, Boli Chen, Stelios Timotheou, and Simos A. Evangelou% <-this % stops a space
\thanks{This work has been supported by the EPSRC Grant EP/N022262/1 and partially funded by the European Union{'}s Horizon 2020 research and innovation programme under grant agreement No 739551 (KIOS CoE) and the Government of the Republic of Cyprus through the Directorate General for European Programmes, Coordination and Development.}
\thanks{X. Pan and S. A. Evangelou are with the Dept. of Electrical and Electronic Engineering, Imperial College London, UK {\tt\small (xiao.pan17@ic.ac.uk, s.evangelou@ic.ac.uk}).}
\thanks{B. Chen is with the Dept. of Electronic and Electrical Engineering, University College London, UK {\tt\small (boli.chen@ucl.ac.uk)}.}
\thanks{S. Timotheou is with the Dept. of Electrical and Computer Engineering and the KIOS Research and Innovation Center of Excellence, University of Cyprus {\tt\small (timotheou.stelios@ucy.ac.cy)}.}}

\begin{document}

\thispagestyle{empty}
\setcounter{page}{0}
\begin{figure*}
\centering
\includegraphics[width=.9\textwidth]{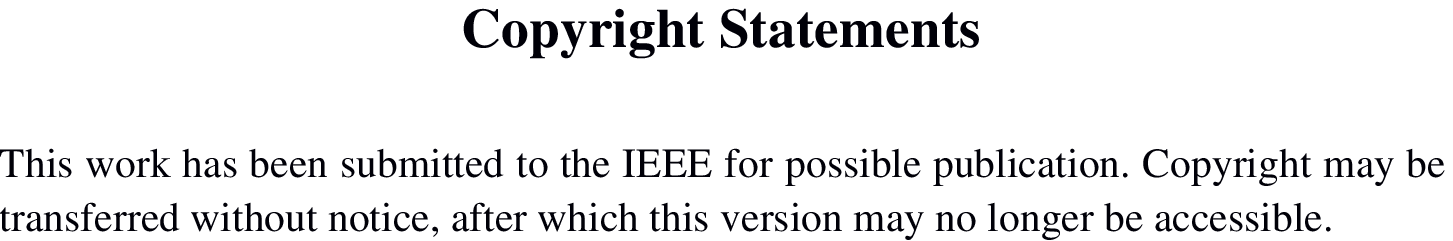}
\end{figure*}

\maketitle

\thispagestyle{fancy}
\chead{This work has been submitted to the IEEE for possible publication. Copyright may be transferred without notice, after which this version may no longer be accessible.}
\rhead{~\thepage~}
\renewcommand{\headrulewidth}{0pt}

\pagestyle{fancy}
\chead{This work has been submitted to the IEEE for possible publication. Copyright may be transferred without notice, after which this version may no longer be accessible.}
\rhead{~\thepage~}
\renewcommand{\headrulewidth}{0pt}

\begin{abstract}
Cooperative vehicle management emerges as a promising solution to improve road traffic safety and efficiency. This paper addresses the speed planning problem for connected and autonomous vehicles (CAVs) at an unsignalized intersection with consideration of turning maneuvers. The problem is approached by a hierarchical centralized coordination scheme that successively optimizes the crossing order and velocity trajectories of a group of vehicles so as to minimize their total energy consumption and travel time required to pass the intersection. For an accurate estimate of the energy consumption of each CAV, the vehicle modeling framework in this paper captures 1) friction losses that affect longitudinal vehicle dynamics, and 2) the powertrain of each CAV in line with a battery-electric architecture. 
It is shown that the underlying optimization problem subject to safety constraints for powertrain operation, cornering and collision avoidance, after convexification and relaxation in some aspects can be formulated as two second-order cone programs, which ensures a rapid solution search and a unique global optimum. 
Simulation case studies are provided showing the tightness of the convex relaxation bounds, the overall effectiveness of the proposed approach, and its advantages over a benchmark solution invoking the widely used first-in-first-out policy. The investigation of Pareto optimal solutions for the two objectives (travel time and energy consumption) highlights the importance of optimizing their trade-off, as small compromises in travel time could produce significant energy savings.
\end{abstract}

%%%%%%%%%%%%%%%%%%%%%%%%%%%%%%%%%%%%%%%%%%%%%%%%%%%%%%%%%%%%%%%%%%%%%%%%%%%%%%%%

\section{Introduction}
Urbanization and the steady increase in vehicle numbers are pushing transportation to its limits, resulting in severe congestion, higher emissions and energy consumption, driver discomfort and safety issues. 
These issues have promoted the development of connected and autonomous vehicles (CAVs), which can potentially mitigate the underlying problems \cite{guanetti2018control}. 
Moreover, there is a growing interest in studying cooperative vehicle management at intersections, which greatly impacts the overall efficiency of road traffic systems, especially in urban areas. 
The performance of traditional traffic light control systems is limited due to the lack of sensing and communication capabilities. 
This incentivizes innovative intersection control schemes, which can leverage advanced vehicular communication systems. A comprehensive overview of recent advancements in intersection management techniques for both signalized and unsignalized intersections is presented in \cite{Chen:tits16}. 
A cooperative control strategy for a signalized intersection coordinates the velocity of each CAV by utilizing signal phasing and timing, which are received via vehicle-to-infrastructure (V2I) communication to minimize the vehicle queuing time \cite{Yang:tits17,Wang:icits17,Kamal:tits15,Wei:icpic14}. The present paper focuses on unsignalized intersections, where the constraints introduced by the traffic lights are removed and therefore they have the potential of further reducing traffic delays and vehicle energy usage~\cite{bian2019reducing,choi2019reservation,vcakija2019autonomous,castiglione2020cooperative}. 

A large number of unsignalized intersection control methods, from centralized to decentralized, have been proposed in the literature \cite{Namazi:ia19,Rios:tits17,Guo:trc2019}. 
A second important categorization of cooperative driving is vehicle prioritization, with the prevalent approaches being:
 1) First-In-First-Out (FIFO), and 2) planning/scheduling-based. The former represents a straightforward and easy-to-implement protocol, whereas the latter also includes the determination of the optimal crossing order subject to safety and throughput requirements \cite{fayazi2018mixed,Fayazi:acc17,Mihaly:ifac2020}. A comparative study of the two strategies is reported in \cite{Meng:18}. It is shown that the crossing sequence sought by a planning-based approach usually leads to more efficient traffic flow than the FIFO policy and the difference can become even more apparent in high-volume traffic situations. However, the computational cost of the planning-based approaches is much higher and it grows exponentially as the number of vehicles increases. 

The foundation of centralized approaches involves a single central controller that determines the velocity trajectories of all the CAVs crossing the intersection. Common centralized approaches are optimization-based with the main objective of either increasing the throughput of an intersection (which can be achieved by minimizing the travel time) \cite{Liu:tvt2020} or reducing the vehicle energy consumption \cite{Murgovski:cdc15}. In particular, \cite{Riegger:itsc16} proposes a centralized model predictive control framework, where the intersection crossing problem is formulated as a convex quadratic program for minimizing energy consumption by following a specific crossing sequence. However, single objective optimization is not sufficient in most cases due to the trade-off between the two goals of minimizing travel time and energy consumption. In particular, minimizing the travel time usually incurs high levels of vehicle energy consumption and vice versa. Recent research effort has focused on the co-optimization of energy consumption and travel time to find the optimal trade-off in \cite{hadjigeorgiou19,chen:ifac2020,CDC2020}, where hierarchical and convex optimization approaches are developed subject to a FIFO policy. Moreover, \cite{Zhao:ITSC18} presents a multi-objective optimization method focusing on coordinating vehicles to improve also driving comfort, in addition to energy and travel time costs. 

There have been numerous other efforts reported in the literature based on decentralized control frameworks, where the velocity of each individual CAV is found by sequentially solving local optimization associated with each vehicle \cite{qian2015decentralized,Medina:tits18,Krajewski:ivs16,Malikopoulos:2018,Campos:17,Wu:19,khoury2019practical}. An analytic optimization method is proposed in \cite{Malikopoulos:2018} to address each local optimization problem subject to a throughput maximization requirement. Reference \cite{Campos:17} presents a sequential optimal control approach that is combined with a computationally efficient scheduling method to maximize throughput. In \cite{Wu:19}, the sequential movement of CAVs is modeled using multi-agent Markov decision processes, while reinforcement learning is employed to find each velocity trajectory. Turning maneuvers are integrated in \cite{Mirheli:trc2019,zhang2017decentralized}, which present distributed coordinated frameworks capable of finding near-optimal solutions. The work in \cite{Mirheli:trc2019} solves the cooperative trajectory planning problem as vehicle-level mixed-integer non-linear programs, whereas the work in \cite{zhang2017decentralized} solves the problem resorting to analytic optimization techniques. More computationally efficient alternatives to optimization-based approaches are heuristic control strategies \cite{zhang2015state,chouhan2018autonomous,Khoury:jits2019}, which however do not have optimality guarantees in most cases. 

A common drawback of all aforementioned methods is that they sacrifice the accuracy of the overall modeling framework to simplify the considered problem. The modeling sacrifice occurs at four levels:
\begin{enumerate}[label=\alph*)]
\item Non-optimal FIFO protocols are usually imposed to streamline the coordination at a signal-free intersection at the cost of coordination controller optimality.
\item Velocity trajectories in the majority of the works are not optimized over the entire intersection area (including the center of the intersection and the immediate vicinity). The control actions stop immediately after the vehicles leave the center of the intersection. Moreover, the speed of CAVs at the center of the intersection is usually assumed constant (non-optimal), which reduces the control complexity but lacks optimality.
\item A lossless linear longitudinal model of each vehicle is commonly utilized, where friction and aerodynamic losses are neglected, and the vehicle acceleration acts as the control input.
\item The somewhat inaccurate metric of $L^2$-norm of the acceleration is considered to represent the energy consumption of each CAV and even without consideration of powertrain losses that may be significant \cite{hadjigeorgiou19,Malikopoulos:2018}.
\end{enumerate}
Such formulations can maintain the problem complexity at a manageable level even for a large number of CAVs, however, the modeling inaccuracy would lead to suboptimal velocity trajectories. Earlier work of the authors~\cite{chen:ifac2020} has proposed a new convex optimization paradigm in space coordinates utilizing a realistic longitudinal vehicle model with nonlinear resistance losses. The work has been extended in \cite{CDC2020} by integrating a battery-electric powertrain system (capable of predicting powertrain losses) to enable more accurate control solutions. The modeling framework employed in \cite{Hult:ecc19} also overcomes the modeling shortcomings b), c) and d) and furthermore it involves turning maneuvers. However, the crossing order in \cite{Hult:ecc19} is predefined rather than being optimized and the coordination problem is formulated as a nonlinear optimization problem, which is computationally demanding.

The present paper expands the work in \cite{CDC2020} to address the autonomous intersection crossing problem for realistic scenarios and with superior optimization efficiency, by introducing and solving an important new combination of modeling features and by enhancing the underlying optimization framework with novel problem formulations and solutions. In more detail, the contributions of the present work are:

\begin{enumerate}
   \item The proposed coordination scheme optimizes in a hierarchical fashion the crossing order and the corresponding velocity trajectories, involving turning maneuvers, within the entire intersection area without invoking restrictive assumptions on a) the crossing order (e.g., FIFO) or b) the actual or average vehicle speed at any point within the intersection or its center, which was the case in many existing works, such as~\cite{hadjigeorgiou19,zhang2017decentralized,Hult:ecc19}. This new and more challenging control paradigm can lead to a further optimized solution.
	 \item Efficient solutions for the formulated coordinated scheme are proposed by suitably relaxing the nonconvex constraints and reformulating both upper (crossing order optimization) and lower (velocity trajectory optimization) level nonconvex OCPs into convex second-order cone programs (SOCPs) using a space-domain modeling approach. 
	\item The optimality of the proposed coordination scheme (with a relaxed SOCP solution) is demonstrated by comparison with a valid lower bounding optimal solution with expanded feasibility. Moreover, the benefit over traditional approaches is investigated by comparison with a benchmark solution that is obtained by following the FIFO policy.
\end{enumerate}

The remainder of this paper is organized as follows. Section~\ref{sec:description} introduces the modeling framework of an autonomous intersection crossing problem, with consideration of the electric powertrain model of each CAV. The associated OCP for energy consumption and travel time minimization is also formulated in Section~\ref{sec:description} , while the hierarchical coordination scheme is introduced in Section~\ref{sec:control} and the SOCP reformulation of the original OCP is performed in Section~\ref{sec:convex}. Simulation results and discussion are presented in Section~\ref{sec:simulation}. Finally, concluding remarks are given in Section~\ref{sec:conclusions}.

%--------------------------------------------------------------------------------------------
\section{Problem Statement}
\label{sec:description}
\subsection{Intersection Model}
\begin{figure}[!ht]
\centering
\includegraphics[width=.8\columnwidth]{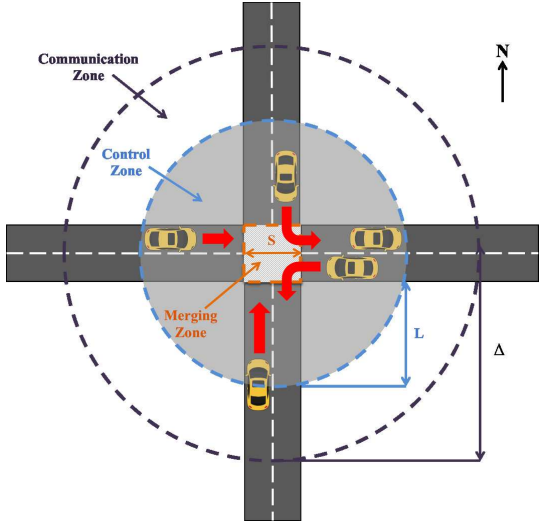}\\[-2ex]
\caption{Autonomous intersection with connected and autonomous vehicles.}
\label{fig:Intersection}
\end{figure}

The present work considers a scenario where multiple homogeneous CAVs approach an unsignalized intersection and need to be regulated to ensure safe, smooth and energy efficient of traffic flow. As illustrated in Fig.~\ref{fig:Intersection}, the intersection contains four branches coming from the north, south, east and west.
For simplicity, it is assumed that all the roads in the intersection are flat with two lanes per perpendicular road (one lane per branch), while lane changes and road slopes are not considered in the present framework.

The centre of the intersection is the \textit{Merging Zone} (MZ), where vehicles merge from different directions, and therefore, it is the area of potential lateral collisions. The area of the MZ is considered to be a square of sides $S$. CAVs are allowed to go straight through, or turn left or right in the MZ, while U-turns are not allowed. The intersection has a central traffic coordinator, called \textit{Intersection Controller} (IC). Vehicles approaching the intersection will first enter a \textit{Communication Zone} (ComZ), where the IC can communicate with each CAV. The radius of the ComZ, $\Delta$, depends on the communication range capability of modern V2I technology. Moving forward towards the intersection, the CAVs will enter from the ComZ to the more compact \textit{Control Zone} (CZ), inside which the motion of each CAV is fully controlled by the IC. The distance from the entry of the CZ to the entry of the MZ is $L$, with $\Delta\!>\!L\!>\!S$ as the sensing range is usually much greater than the physical length of the MZ and the CZ.

At each time, only the vehicles that are running within the annulus between the boundaries of ComZ and CZ will be considered for trajectory planning by the IC. Let us denote $N \!\in\! \mathbb{N}_{>0}$ the total number of CAVs arriving at the ComZ and remaining inside the annulus within a fixed time-interval $T\!\in\! \mathbb{R}_{>0}$, and $\mathcal{N}\!=\!\{1,2,\ldots,N\}\!\in\!\mathbb{Z}^{N}$ the set to designate the agreed order in which the vehicles will cross the intersection. The determination of the crossing order will be elaborated in Section~\ref{sec:control} as part of the coordination scheme. Furthermore, all CAVs are considered as identical battery electric vehicles (BEVs) with vehicle lengths equal to $l,\,(l\!<\!S)$. These assumptions are relatively straightforward to relax as long as the vehicle lengths and powertrain models are known to the central coordinator. The control target is to optimize the total electric energy and time consumption of all $N$ CAVs by determining their speed trajectories (in a centralized manner) from the entry of the CZ to the exit point of the CZ, which depends on the turning decision of a CAV made at the MZ. In the subsequent part of this paper, the decision of the $i$th CAV at the intersection is denoted by $d_i \in \{-1,0,1\}$, wherein $d_i \!=\! 0$ indicates going straight, while $d_i \!=\! -1$ and $d_i \!=\! 1$ indicate a left turn and a right turn, respectively.

The gross motion of CAVs is modeled by using the single-track, non-holonomic vehicle model \cite{chen:2019}. In this context, the longitudinal dynamics are described by the following differential equation:
\begin{equation}\label{eq:sysv}
    \frac{d}{dt} v_i(t) = \frac{F_{w,i}(t) - F_r - F_{d,i}(t)}{m},\,\,i\!\in\! \mathcal{N},
\end{equation}
where $v_i(t)$ is the linear (forward) velocity of the $i$th CAV, $m$ is the vehicle mass, $F_{w,i}(t)$ is the powertrain driving or braking force acting on the wheels, while $F_r\!=\!f_r m g$ and $F_{d,i}(t)\!=\!f_dv_i^2(t)$ are the resistance forces of rolling and air drag, respectively, with $f_r$ and $f_d$ the coefficients of rolling and air drag resistance. Moreover, $F_{w,i}(t)$ can be broken down into two separate control inputs, the powertrain driving force $F_{t,i}(t)$ and the mechanical braking force $F_{b,i}<0$, such that 
\begin{equation}\label{eq:fwi}
F_{w,i}(t) = F_{t,i}(t)+F_{b,i}(t).	
\end{equation}
The main characteristic parameters of the vehicle model are summarized in Table~\ref{tab:vehicledata}. 
\begin{table}[!ht]
\centering 
\caption{Electric Vehicle Model Parameters}
\label{tab:vehicledata} 
\begin{tabular*}{1\columnwidth}{l @{\extracolsep{\fill}} cl}
\hline
\hline
 symbol & value & description\\
 \hline
 $m$ & 1200\,kg & vehicle mass \\
 $r_w$ & 0.3\,m & wheel radius\\
 $g_r$ & 3.5 & transmission gear ratio\\
 $f_r$ & $0.01\,$ & rolling resistance coefficient\\
 $f_d$ & $0.47\,$ & air drag resistance coefficient\\
 $v_{\min}$    &$0.1$\,m/s& minimum velocity \\
 $v_{\max}^f$    &$15$\,m/s& maximum forward velocity \\
 $a_{\min}$  &$-6.5\,\text{m/s}^2$ &   vehicle maximum deceleration  \\
 $l$ &  $4\,\text{m}$ & vehicle length \\
\hline
\hline
\end{tabular*}
\end{table}

It is considered that the traffic follows the \textit{left-hand driving system} and all vehicles travel along the centerline of their lane. The path of a turning CAV at the intersection can be modeled as a $90^{\circ}$ arc with turning radii $R_l=S/4$ for the left turn and $R_r=3S/4$ for the right turn. As such, the driving distance of CAV $i$ inside the MZ is:
\begin{equation}
\delta(d_i)  =
\left\{\begin{array}{ll}
S,  & d_i=0 ,\\
\displaystyle \frac{1}{8} \pi S, &  d_i=-1,\\[2ex]
\displaystyle \frac{3}{8}\pi S,  & d_i=1.
\end{array}\right.
\end{equation}
Considering that the mission starts when the front of the vehicle enters the CZ and ends when its front reaches the exit of the CZ, the total travel distance of each CAV within the CZ is $2L+\delta(d_i)$.

\salt 

Instead of formulating the problem in the time domain, which is the common approach used in existing works, this paper defines the intersection problem in the space domain, which is beneficial for acquiring the vehicle arrival times at any given position inside the CZ. This fact turns out to be very useful in establishing time-dependent constraints and obtaining the travel time of each CAV required to cross the intersection. The latter aspect greatly facilitates the joint optimization formulation of energy and travel time costs rather than requiring additional steps to tackle the free end-time problem that appears if the problem is formulated in the time domain \cite{hadjigeorgiou19}. The use of the space domain also allows the overall intersection management problem to be formulated as an SOCP, which offers a rapid and unique solution search as compared to nonconvex counterparts, as will be shown later in Section~\ref{sec:convex}.

Let $s$ denote the variable of traveled distance. By changing the independent variable $t$ to $s$ via $\frac{d}{ds} = \frac{1}{v_i}\frac{d}{dt}$, the differential equation \eqref{eq:sysv} describing the vehicle longitudinal dynamics can be rewritten as:
\begin{equation}\label{eq:dynamic_dt}
    \frac{d}{ds} v_i(s) = \frac{F_{w,i}(s)\!-\!F_r\!-\! F_{d,i}(s)}{m v_i(s)},\,\, i\!\in\! \mathcal{N},
\end{equation}
for all $s\!\in\! [0,2L\!+\!\delta(d_i)]$.
As the model is formulated in the space domain, the individual travel time $t_i$ of each CAV is introduced as a system state, with its dynamics expressed by:
\begin{equation}\label{eq:syst}
\frac{d}{ds} t_i(s) = \frac{1}{v_i(s)}, \, i\!\in\! \mathcal{N},    
\end{equation}
from which the required travel time of each CAV to traverse the CZ can be expressed as,
\begin{equation}\label{eq:Jti}
J_{t,i}=t_i(2L+\delta(d_i))\!-\!t_i(0),
\end{equation}
where $t_i(0)$ is the entry time of CAV $i$ in the CZ.
The velocity of each CAV is constrained by:
\begin{equation}
v_{\min} \leq \,v_i\, \leq v_{\max}(s),
\label{eq:vbound}  
\end{equation}
where $v_{\min}$ is set to a sufficiently small positive constant to avoid singularity issues in \eqref{eq:syst} without sacrificing the generality of the formulation, and $v_{\max}(s)$ is set for safety and comfort purposes and depends on the vehicle turning decision and the position at the intersection. More specifically, the upper speed limit $v_{\max}(s)$ is defined as a piecewise constant function, which is $v_{\max}(s)=v_{\max}^f,\,\forall s \in [0,L) \cup (L+\delta(d_i),2L+\delta(d_i)]$ whereas for $s \!\in\! [L,L+\delta(d_i)]$ in the MZ, it follows: 
\begin{equation}
v_{\max}(s)  =
\left\{\begin{array}{lll}
%v_{\max}^f,  &\text{going straight} ,\\
v_{\max}^f,  &\text{straight traveling} ,\\
v_{\max}^l, &  \text{left turn},\\
v_{\max}^r,  & \text{right turn},
\end{array}\right.
\end{equation}
with $v_{\max}^l\!<\!v_{\max}^r\!<\!v_{\max}^f$. 
%-----------------------------------------------------------------
In this work, speed limits for cornering, $v_{\max}^l,\,v_{\max}^r$, are estimated by utilizing the acceleration diamond \cite{chen:2019} that represents a constraint for ordinary drivers on longitudinal and lateral accelerations for comfortable driving (away from the limits of tire adherence on the road):
\begin{equation}\label{eq:comfort}
%  \left| \displaystyle\frac{(F_{w,i} - F_r - F_{d,i})/m} {a_{x,max}} \right|
    \left| \displaystyle\frac{F_{w,i}/m} {a_{x,max}} \right|
  + \left| \displaystyle\frac{v_i\Omega_i}{a_{y,max}} \right| \le 1\,,
\end{equation}
where $\Omega_i$ is the vehicle yaw rate, defined by,
\begin{equation}\label{eq:omega}
\Omega_i = \left\{
 \begin{array}{ll}
   v_i/R_l ,   &  \text{left turn},\\
   v_i/R_r ,   &  \text{right turn},
 \end{array}  \right.
\end{equation}
and the physical limits of the longitudinal and lateral accelerations, $a_{x,max}$ and $a_{y,max}$ respectively, are chosen to be the nominal gravitational acceleration $g$. Note that ${F_{w,i}}/{m}$ represents a good approximation of the longitudinal acceleration (by ignoring the subtle influence of resistive forces $F_r$ and $F_{d,i}$) and $v_i\Omega_i$ corresponds to the lateral acceleration. The acceleration diamond conforming to \eqref{eq:comfort} is illustrated in Fig.~\ref{fig:tirelimit}. As it can be seen, the longitudinal deceleration/acceleration of each BEV is bounded by $[F_{w,\min}/m,\,F_{w,\max}/m]$, where $F_{w,\max}$ and $F_{w,\min}$ represent the maximum and minimum available driving force acting on the wheels, respectively. 
Note that the value of $F_{w,\max}$ will be defined later in Section~\ref{sec:powertrain} on the basis of the  motor efficiency map and associated motor torque limits shown in Fig.~\ref{fig:motormap}, and $F_{w,\min}$ is chosen such that $a_{\min}\!=\!F_{w,\min}/m$ be the peak deceleration for braking, which provides a reasonable margin to the limit of tire adhesion. 
In order to streamline the modeling framework, under cornering the admissible zone for operation is taken as the rectangular dark gray area shown in Fig.~\ref{fig:tirelimit}, which corresponds to the uniformed longitudinal acceleration limits, $[-F_{w,\max}/m,\,F_{w,\max}/m]$. 
Operation within this region can be provided entirely by the powertrain driving force $F_{t,i}$, where any amount of braking can be fully regenerated, and which is compatible with the target of eco-driving sought in the present work.  
Note that under straight running conditions, it is further allowed to operate a CAV in the light gray area below the mentioned dark gray area, however, in that case additional mechanical braking, $F_{b,i}$, would be required. 
As the feasible dark gray region also defines the maximum permissible lateral acceleration (i.e., $v\Omega$), which is realized at the four corners of the rectangular dark gray area, the maximum permissible speed for left and right turns, respectively, can consequently be derived from \eqref{eq:comfort} and \eqref{eq:omega} as follows:
\begin{figure}[!ht]
\centering
\includegraphics[width=.72\columnwidth]{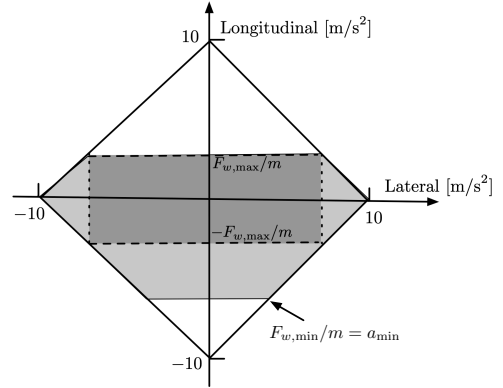}\\[-2ex]
\caption{Theoretical driving comfort limits represented by an acceleration diamond \cite{chen:2019} (thick solid lines) and the theoretical performance envelope of the BEV (light gray area). 
For simplicity of implementation, a practical performance region of the BEV (dark gray area) under cornering conditions is assumed to be enveloped by longitudinal acceleration saturated at $\pm F_{w,\max}/m$ (horizontal dashed lines) and conservative lateral acceleration limits (vertical dotted lines). $F_{w,\max}$ is calculated based on the BEV motor torque limits shown in Fig~\ref{fig:motormap} and $a_{\min}$ is given in Table~\ref{tab:vehicledata}.} 
\label{fig:tirelimit}
\end{figure}
  
\begin{subequations}
\begin{align}
v_{\max}^l = \sqrt{\left(1-\frac{F_{w,\max}/m} {a_{x,\max}}\right){a_{y,\max}}R_l}%=5m/s
\,,\\
v_{\max}^r =  \sqrt{\left(1-\frac{F_{w,\max}/m} {a_{x,\max}}\right){a_{y,\max}}R_r}%=8.7m/s
\,.
\end{align}
\label{eq:turningspeedlimits}
\end{subequations}

Additional constraints, the collision avoidance constraints, are designed to address further safety challenges of the unsignalized intersection coordination problem. Note that, given an arbitrary CAV $i$ and $s \!\in\! [0,L]$, any CAV $j\!\in\! \mathcal{N}, \, j \!>\! i$ must belong to one of the following four subsets of $\mathcal{N}$:
1) $\mathcal{C}_i$ collects vehicles traveling in the same direction as the $i$th vehicle; 2) $\mathcal{O}_i$ collects vehicles traveling in the opposite direction to the $i$th vehicle; 3) $\mathcal{P}_{L,i}$ and $\mathcal{P}_{R,i}$ collect vehicles traveling in the perpendicular directions to the $i$th vehicle from its left and right hand sides, respectively. 

Assuming CAV $i$ is immediately ahead of CAV $k$ with $k\in\mathcal{C}_i$, the following constraint on the time gap,  $t_k(s)\!-\!t_i(s+l)$,  is imposed to ensure the absence of rear-end collisions 
\begin{equation}\label{eq:TTC}
  t_k(s)\!-\!t_i(s+l) > \max\left(\frac{v_k(s)-v_i(s+l)}{|a_{\min}|}, t_{\delta}\right),
\end{equation}
where the first term in the max function of the inequality represents the time-to-collision~\cite{liu2019comparison}, and $t_{\delta}$ is a small time constant to enforce a safety margin invariably. 

The rear-end collision avoidance constraint \eqref{eq:TTC} is enforced over the entire space horizon $[0,2L+\delta(d_i)]$ if $d_k\!=\!d_i$. 
However, when $d_k \ne d_i$, \eqref{eq:TTC} is needed only for $s\in [0,L]$ whereas for the rest of the mission, $s \in [L,2L+\delta(d_i)]$, only the following constraint is imposed:
\begin{equation}\label{eq:TTC2}
t_i(L+\delta(d_i)+l)\leq t_k(L),\,\, k\in\mathcal{C}_i ,
\end{equation}
which prevents rear-end collisions between CAVs $i$ and $k$ within the MZ as it does not allow them to travel inside the MZ at the same time. 

\begin{figure*}%[htb!]
\centering
\subfigure[$\mathcal{O}_{i}^r \cup \mathcal{P}_{L,i}\cup \mathcal{P}_{R,i}^s\cup \mathcal{P}_{R,i}^r, \,\,\text{if}\,\, d_i \!=\! 0.$]{
\centering
\includegraphics[width=.25\textwidth]{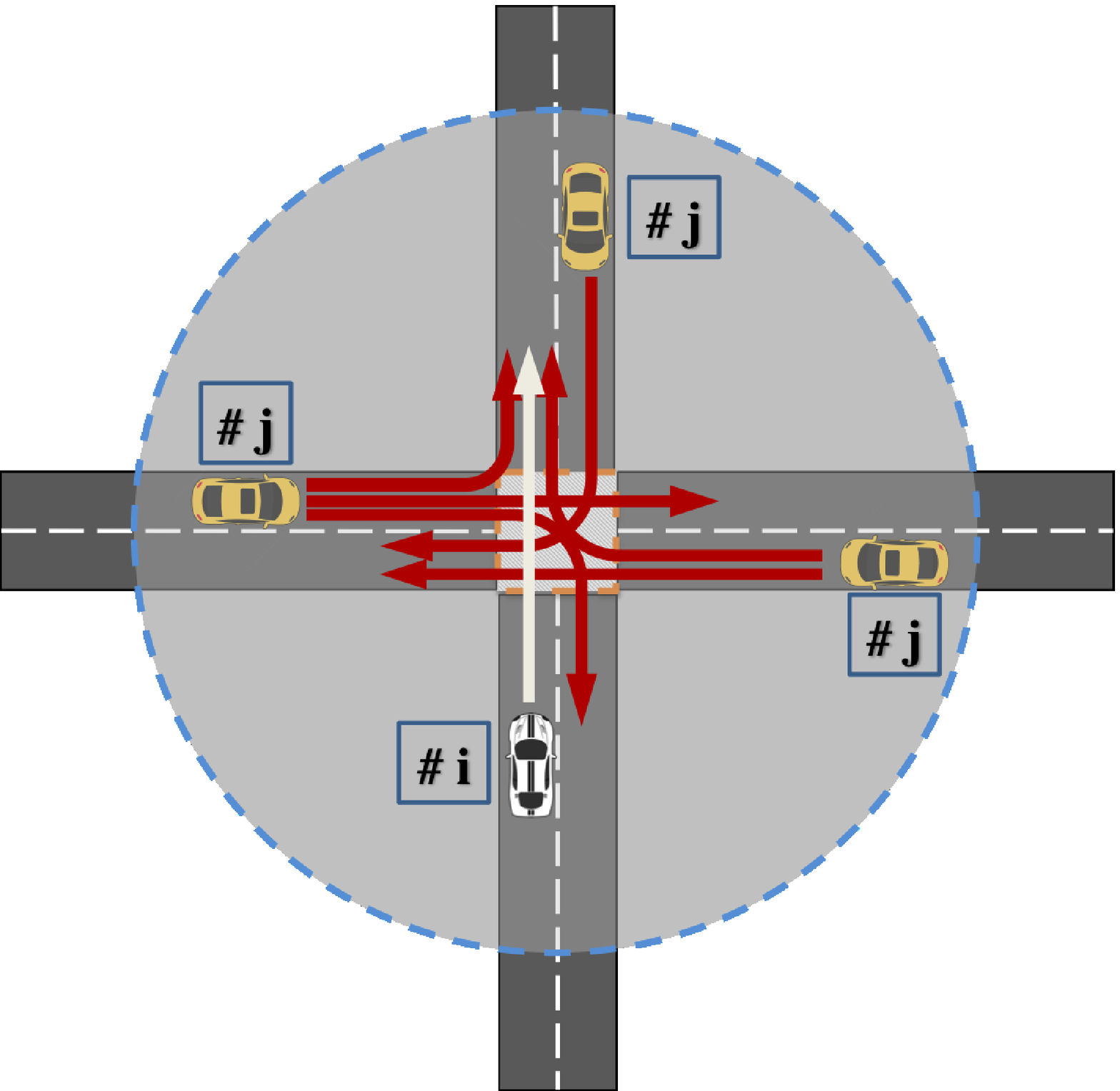}
}
\hspace{.8cm}
\subfigure[$\mathcal{O}_{i}^r \cup \mathcal{P}_{R,i}^s\cup \mathcal{P}_{R,i}^r,\,\,\text{if}\,\, d_i \!=\! -1.$]{
\centering
\includegraphics[width=.25\textwidth]{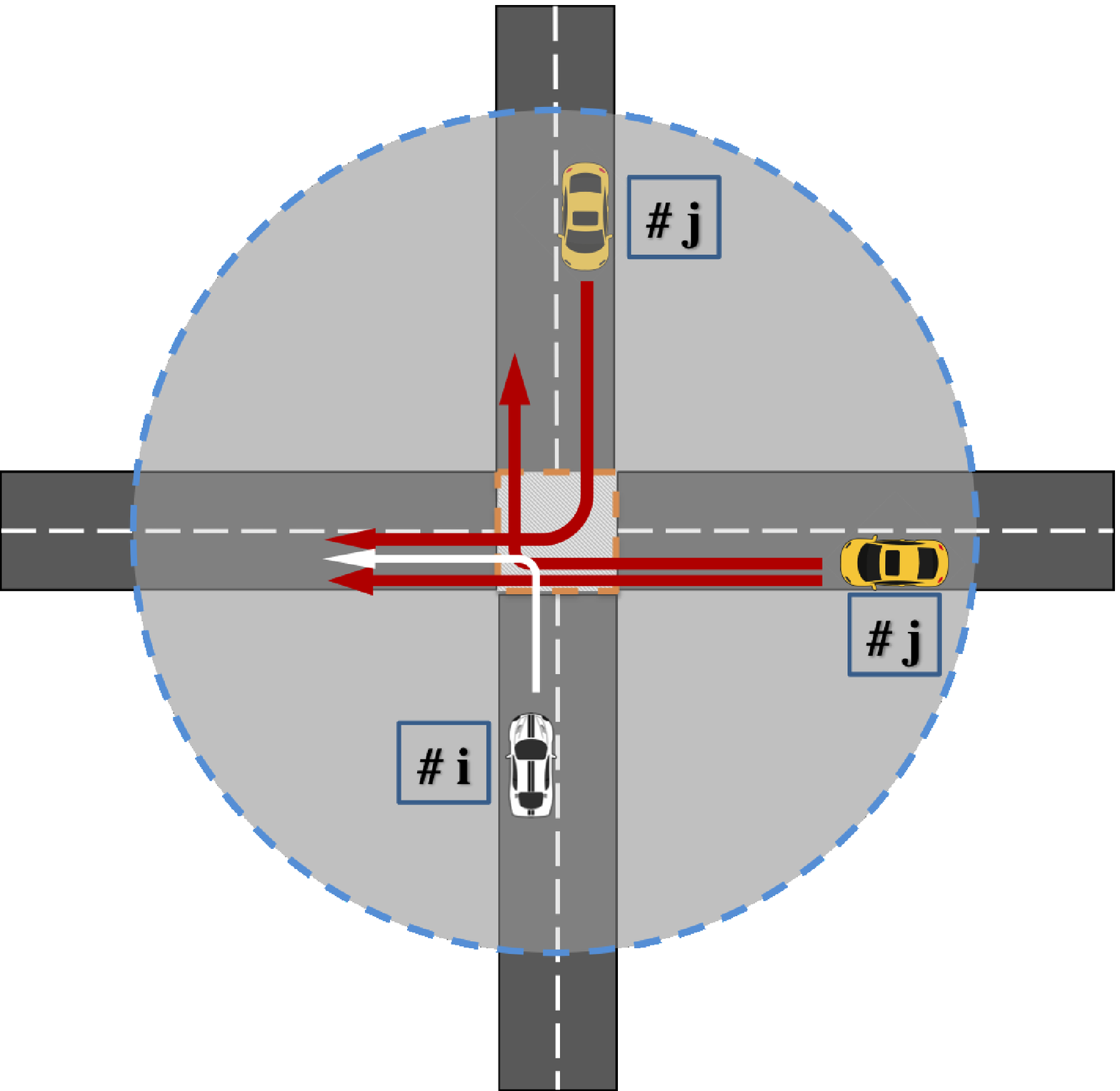}
}
\hspace{.8cm}
\subfigure[$\mathcal{O}_{i} \cup \mathcal{P}_{L,i}\cup \mathcal{P}_{R,i}^s\cup \mathcal{P}_{R,i}^r,\,\,\text{if}\,\, d_i \!=\! 1.$]{
\centering
\includegraphics[width=.25\textwidth]{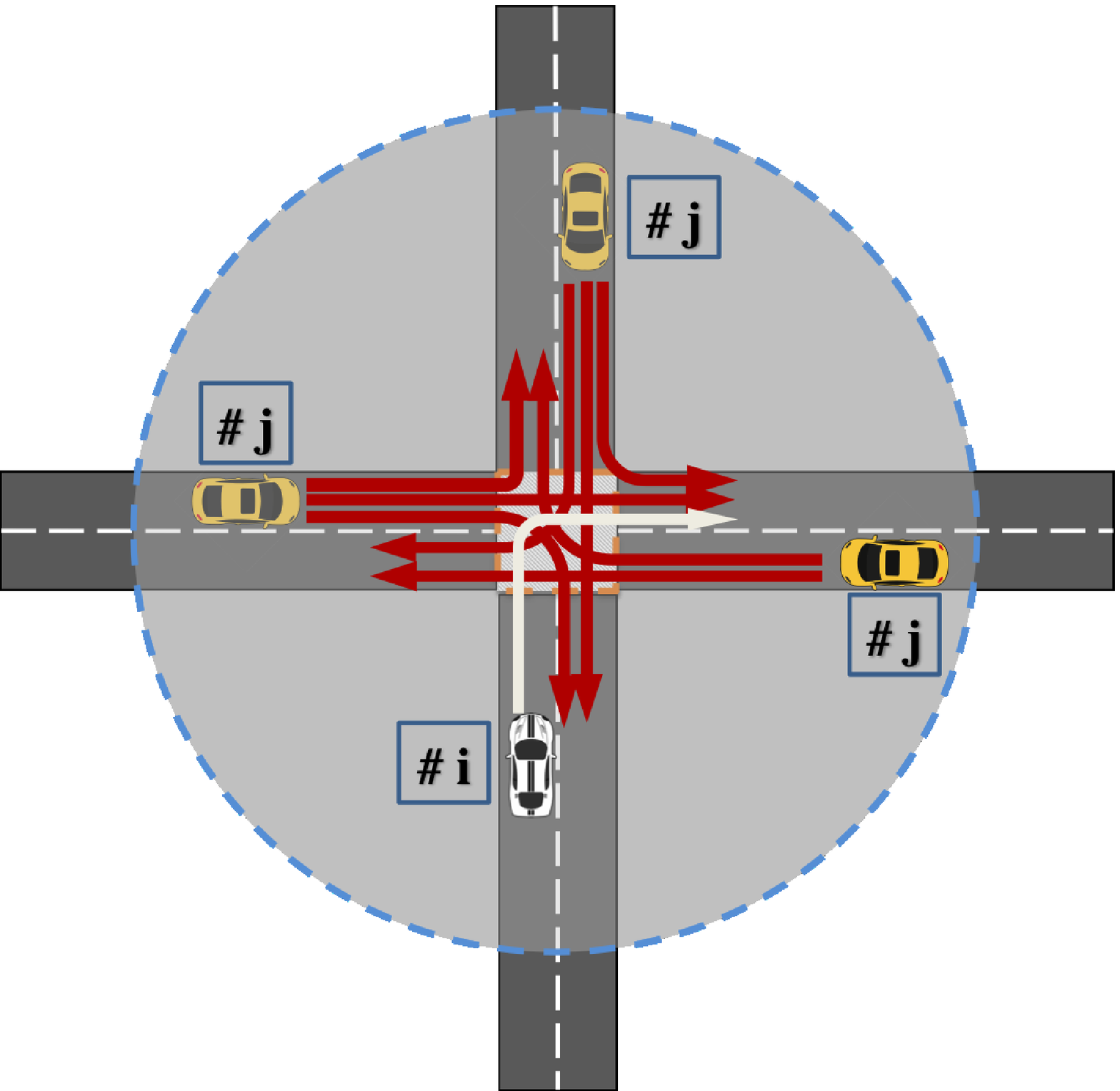}
}\\[-2ex]
\caption{Illustration of potential collisions \eqref{eq:sideconstraint1} between CAVs $i$ (white vehicle) and $j$ (yellow vehicles) at the MZ with $j>i$.}
\label{fig:collision}
\end{figure*}

For any CAV $j \notin \mathcal{C}_i, j>i$, it is clear that a collision can only arise after they enter the MZ, for $s \in [L,2L+\delta(d_i)]$. In order to discuss collision threats (including side and rear-end collisions) in this context, let us further define subsets of $\mathcal{O}_i$, $\mathcal{P}_{L,i}$ and $\mathcal{P}_{R,i}$ in relation to the intention $d_j$ of vehicle $j$, as follows:
\begin{align}
& \mathcal{O}_i = \mathcal{O}_{i}^l\cup\mathcal{O}_{i}^s\cup\mathcal{O}_{i}^r,   \\
& \mathcal{P}_{L,i} = \mathcal{P}_{L,i}^l\cup\mathcal{P}_{L,i}^s\cup\mathcal{P}_{L,i}^r,  \\
&\mathcal{P}_{R,i} = \mathcal{P}_{R,i}^l\cup\mathcal{P}_{R,i}^s\cup\mathcal{P}_{R,i}^r,
\end{align}
where the superscripts $l,s,r$ denote left turn, going straight and right turn of the $j$th vehicle, respectively. 
Potential collisions between CAVs $i$ and $j$ may occur if CAV $j$ belongs to:
\begin{equation} \label{eq:sideconstraint1}
\mathcal{L}_i\triangleq
\left\{
\begin{array}{ll}
  \mathcal{O}_{i}^r \cup \mathcal{P}_{L,i}\cup \mathcal{P}_{R,i}^s\cup \mathcal{P}_{R,i}^r,& \text{if}\,\, d_i \!=\! 0, \\
 \mathcal{O}_{i}^r \cup \mathcal{P}_{R,i}^s\cup \mathcal{P}_{R,i}^r,&\text{if}\,\, d_i \!=\! -1, \\
 \mathcal{O}_{i} \cup \mathcal{P}_{L,i}\cup \mathcal{P}_{R,i}^s\cup \mathcal{P}_{R,i}^r,&\text{if}\,\, d_i \!=\! 1,
\end{array} \right.   
\end{equation}
which is influenced by the decision $d_i$ of the lead CAV $i$ at the MZ. An illustration of these collision models is given in Fig.~\ref{fig:collision}.
To prevent such collisions inside the MZ between vehicles $i$ and  $j\!\in\!\mathcal{L}_i$, the IC simply follows the rule that the $j$th CAV enters the MZ only after CAV $i$ has left the MZ, and once again \eqref{eq:TTC2} is applicable. If vehicle $j$ further belongs to $\mathcal{D}_i$ that is a subset of $\mathcal{L}_i$, defined by:
\begin{equation}\label{eq:rearend_after}
\mathcal{D}_i\triangleq
\left\{
\begin{array}{ll}
  \mathcal{P}_{L,i}^l\cup \mathcal{P}_{R,i}^r,& \text{if}\,\, d_i \!=\! 0, \\
 \mathcal{O}_{i}^r \cup \mathcal{P}_{R,i}^s,&\text{if}\,\, d_i \!=\! -1, \\
 \mathcal{O}_{i}^l \cup \mathcal{P}_{L,i}^s,&\text{if}\,\, d_i \!=\! 1,
\end{array} \right.   
\end{equation}
both CAVs $i$ and $j$ merge into the same lanes after the MZ. As such, the rear-end collision avoidance constraint \eqref{eq:TTC} is once again required for $s \!\in\! [L+\delta({d_i}),2L+\delta({d_i})]$ so as to avoid potential rear-end collision between these vehicles. Otherwise, when $j\!\in\!\mathcal{L}_i \backslash \mathcal{D}_i$, then both vehicles travel towards different directions after leaving the MZ, which excludes the possibility of collisions outside the MZ. The above characterization of the potential collisions \eqref{eq:rearend_after} after the MZ is a novelty of the present work and extremely important when turns are allowed, as vehicles may travel in the same direction after leaving the MZ, but with a noticeable difference in speed, which could lead to a rear-end collision immediately after the MZ; for example, a left-turning vehicle traveling at low speed followed by a straight driving vehicle traveling at high speed in a perpendicular direction. The treatment of these scenarios has not been addressed in the literature as it is only permitted with the extended control horizon that spans the whole CZ, including the section after the MZ, which is one of the important contributions of the present paper.

Finally, for any CAV $h \notin \mathcal{L}_i \cup \mathcal{C}_i$, there is no interference between CAV $h$ and $i$ inside the MZ. Hence, only the following constraint is required: 
\begin{equation}\label{eq:opposite}
t_i(L+\delta(d_i)) \leq t_h(L+\delta(d_i)),\, \,\,h>i.
\end{equation}
As opposed to the more restrictive constraint \eqref{eq:TTC2}, constraint \eqref{eq:opposite} allows multiple CAVs in and to exit the MZ at the same time.
\subsection{Powertrain System and Energy Consumption Model}\label{sec:powertrain}
To formulate the vehicle energy consumption, it is essential to include the tank-to-wheel energy path, which depends on the powertrain of the vehicle. As the CAVs are all battery electric vehicles, their energy consumption can be evaluated by their battery energy cost. 

\begin{figure}[htb!]
\centering
\includegraphics[width=.9\columnwidth]{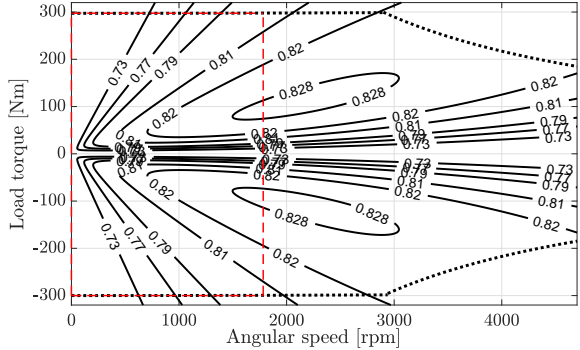}\\[-2ex]
\caption{Efficiency map of the electric motor (positive torque indicates battery discharging and negative torque represents battery charging) and operational bounds (dotted lines) for the reversible motor. The area surrounded by red dashed lines denotes the operational region for the feasible vehicle speed specified by \eqref{eq:vbound}.}
\label{fig:motormap}
\end{figure}

The powertrain connected to the battery contains a DC/DC converter, an electric motor and a transmission set, where both the converter and the transmission are simply modeled by constant efficiency factors and the efficiency of the motor is modeled as a static efficiency map \cite{chen:2019}. Fig.~\ref{fig:motormap} shows the joint efficiency map of all three components.
According to established literature, the input power (electric side) of the motor, $P_{b,i}$, can be represented as a quadratic function of motor force, $F_{t,i}$, and vehicle speed (equivalent to motor torque and angular speed respectively), given by~\cite{Lopes2019EnergySF}:
\begin{equation}\label{eq:Pm2}
\begin{aligned}
P_{b,i}=b_1 F_{t,i}^2 + b_2 F_{t,i}v_i,\,\,i\!\in\! \mathcal{N},
\end{aligned}
\end{equation}
where $b_1$ and $b_2$ are fitting parameters. The motor torque is constrained by speed dependent limits $T_{i} \!\in\! [T_{\min}(\omega_i), T_{\max}(\omega_i)]$ such that the operational limits of 
$F_{t,i}$ become,
\begin{equation}\label{eq:forwardforcebound}
\frac{g_r}{r_w}T_{\min} \leq F_{t,i} \leq \frac{g_r}{r_w}T_{\max}\,,
\end{equation}
with $g_r$ and $r_w$ the transmission gear ratio and the wheel radius, respectively. From Fig.~\ref{fig:motormap}, the motor torque limits $T_{\max}$ and $T_{\min}$ are constant until the angular speed reaches approximately 2800 rpm, which corresponds to a forward speed at 25~m/s. However, it is reasonable to impose a much lower speed limit for vehicles approaching an intersection. Considering the speed limit $v_{\max}^f$ for straight running (maximum possible speed within CZ) given in Table~\ref{tab:vehicledata}, constant motor toque limits ($\pm 300$~Nm) can be applied, as illustrated in Fig.~\ref{fig:motormap}. 
Overall, the physical limits of the force on the wheels from the powertrain, $F_{w,i}$ (see \eqref{eq:fwi}), can be expressed as:
\begin{equation}\label{eq:forcebound}
m a_{\min} \leq F_{w,i} \leq \frac{g_r}{r_w}T_{\max}\left(=F_{w,\max}\right)\,.
\end{equation}

Battery power is a nonlinear function of $P_{b,i}$ due to the internal resistive losses, which are insignificant compared to other powertrain losses~\cite{sciarretta2015optimal}. Hence, it is reasonable to neglect the influence of the battery internal resistance and to approximate the battery power by $P_{b,i}$, such that the battery energy usage of a single CAV can be found by integrating $P_{b,i}$ in the space domain as follows:
\begin{equation}\label{eq:battery_usage}
J_{b,i} =  \int_{0}^{2L+\delta(d_i)}\frac{P_{b,i}(s)}{v_i(s)}\,ds.
\end{equation}

The following assumptions are imposed to complete the modeling framework described above.

\begin{assumption}\label{ass:delay}
	All the vehicle information (e.g., position, velocity, acceleration) can be measured precisely, and the data can be transferred between each CAV and the IC without errors and delays.
\end{assumption} 

\begin{assumption}\label{ass:perfectfollow}
After entering the CZ, all CAVs are capable of precisely following the trajectories provided by the IC with no deviations, e.g. due to modeling and measurement errors.
\end{assumption} 

\begin{assumption}\label{ass:initialprediction}
	The decision of each vehicle on whether a turn is to be made at the MZ and their arrival time and speed at the CZ are known upon its entry at the ComZ.
\end{assumption}

\begin{assumption}\label{ass:initial}
For each CAV $i$, constraints \eqref{eq:vbound}, \eqref{eq:TTC} and \eqref{eq:forcebound} are inactive at $t_i(0)$.
\end{assumption}

\begin{assumption}\label{ass:terminal}
All the CAVs leave the CZ at the same terminal speed $\bar{v} \in [v_{\min},v_{\max}]$, such that
\begin{equation}\label{eq:terminalv}
  v_i(2L+\delta(d_i)) = \bar{v} ,\,\,\forall i\in \mathcal{N}
\end{equation}
\end{assumption}
 Assumption \ref{ass:delay} - \ref{ass:perfectfollow} may be not valid for practical vehicular networks. On that occasion, it can be relaxed by using a worst-case analysis as long as the measurement and communication uncertainties are bounded. More conservative safety constraints can be formulated by incorporating the upper bounds on the uncertainties (e.g., measurement noise, communication delay, etc). Assumption \ref{ass:initialprediction} is essential to implement the centralized optimal control scheme. The feasibility of this assumption can be justified if CAVs are managed to enter the CZ at a predefined speed and time. Relaxation of the assumption can be accomplished by using a model predictive control framework. Assumption~\ref{ass:initial} is needed to ensure that all CAVs arriving at the CZ have feasible initial states and initial control inputs. The Assumption~\ref{ass:terminal} is intended to streamline the framework and to allow solutions in different scenarios to be compared. Further relaxation of \eqref{eq:terminalv} (non-uniform $v_i(2L+\delta(d_i))$) can be made easily, if necessary. 
 
\subsection{Optimal Control Problem Formulation}\label{subsec:ocp}
The central IC aims to find the crossing order and speed trajectories of all CAVs which minimizes a weighted sum of the aggregate battery electric energy, $J_{b,i}$ as defined in \eqref{eq:battery_usage}, and traveling time, $J_{t,i}$ as defined in \eqref{eq:Jti}, subjected to the aforementioned constraints related to vehicle  physical limits and safety regulations. For a specific crossing order $\mathcal{N}$, this optimal control problem can be formulated as:
\begin{problem}\label{prob:ocp}
\begin{subequations}\label{eq:OCP}
\begin{align}
   %\text{minimize:}\quad
    & \hspace{-5mm}\mathop {\min}\limits_{\mathbf{u}} \hspace{4mm} J(\mathbf{x},\mathbf{u})=\sum_{i=1}^{N} W_1 J_{t,i} +W_2 J_{b,i} 
     \label{eq:J} \\
    \textbf{s.t.}:\quad
    & \frac{d}{ds}\mathbf{x} = \mathbf{f} \left(\mathbf{x},\mathbf{u},s\right) \label{eq:state-space}\\
    & {\bm \psi} \left(\mathbf{x},\mathbf{u},s\right) \le \mathbf{0}    \label{eq:constraints}\\
    & {\bm \phi}  \left(\mathbf{x}(0),\mathbf{x}(2L+\delta(d_i))\right) = \mathbf{0}, \label{eq:BC}
\end{align}
\end{subequations}
\end{problem}
where  
\begin{align}\label{eq:control-state-variable}
 & {\bf {x}} = [v_1,v_2,\ldots,v_N,t_1,t_2,\ldots,t_N]^{\top}, \nonumber \\
 & {\bf {u}} = [F_{t,1},F_{t,2},\ldots,F_{t,N},F_{b,1},F_{b,2},\ldots,F_{b,N}]^{\top}, 
\end{align}
represent the system state and control variables, respectively. $W_1,\,W_2$ are the weighting factors tuned to balance the trade-off between the two objectives. The state vector $\mathbf{x}$ evolves according to the dynamic system \eqref{eq:state-space} that encompasses \eqref{eq:dynamic_dt} and \eqref{eq:syst}. The inequality constraints \eqref{eq:vbound}, \eqref{eq:TTC}, \eqref{eq:TTC2}, \eqref{eq:opposite} and \eqref{eq:forwardforcebound}-\eqref{eq:forcebound} are taken into account by \eqref{eq:constraints}. The boundary conditions \eqref{eq:BC} specify the initial and terminal conditions of the states, which fulfil the conditions imposed in Assumptions~\ref{ass:initial} and \ref{ass:terminal}. 
  To find the global optimum, \cref{prob:ocp} has to be solved for all possible crossing orders. For a scenario with $N$ vehicles, there are potentially
$N!$ different orders under which the vehicles can cross the intersection. Therefore, the problem becomes intractable for practical problem sizes.

\section{The Hierarchical Centralized Coordination Scheme}\label{sec:control}

\begin{figure}[htb!]
\centering
\includegraphics[width=.95\columnwidth]{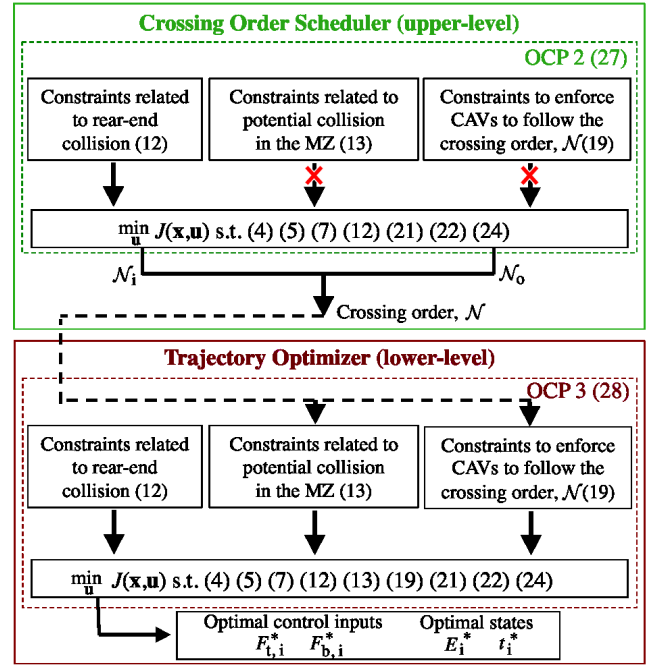}\\[-2ex]
\caption{The hierarchical centralized coordination scheme. $\mathcal{N}_i$ and $\mathcal{N}_o$ represent the resulting CAV orders solved by \cref{prob:ocp1} based on the entry and exit times at MZ, respectively, defined in \cref{subsec:Scheduler}.}
\label{fig:Hierarchical_Scheme}
\end{figure}

In this section, we present an efficient control-based coordination strategy, which circumvents the complexity issue introduced in Section~\ref{subsec:ocp}, and yet it provides close-to-optimal solutions. As shown in Fig.~\ref{fig:Hierarchical_Scheme}, the control framework is composed of two levels, a crossing order scheduler and a trajectory optimizer, deployed in a hierarchical manner.
In particular, the upper-level scheduler determines an optimal crossing order $\mathcal{N}$ whereas the lower-level optimizer solves the optimal trajectories of all CAVs so that collision avoidance is guaranteed for the given order. Each scheme involves solving an individual coordination problem in a centralized fashion. Next, we specify the two steps of our approach.

\subsection{Crossing Order Scheduler}\label{subsec:Scheduler} 
Considering the single lane road depicted in Fig.~\ref{fig:Intersection}, CAVs approaching the intersection from the same direction will enter and leave the MZ in the same order they arrive at the CZ. 
Therefore, the major challenge stems from the collision avoidance and prioritizing constraints for CAVs merging from different directions, i.e., \eqref{eq:TTC2} and \eqref{eq:opposite}, which depend on the crossing order. To determine an optimal crossing order without invoking an exhaustive search, we define a virtual coordination problem in the form of \cref{prob:ocp}:
\begin{problem}\label{prob:ocp1}
\begin{subequations}\label{eq:OCP1}
\begin{align}
   %\text{minimize:}\quad
    & \hspace{-5mm}\mathop {\min}\limits_{\mathbf{u}} \hspace{4mm}  J(\mathbf{x},\mathbf{u}) 
     \label{eq:J1} \\
    \textbf{s.t.}:\quad
    & \eqref{eq:dynamic_dt}, \eqref{eq:syst}, \eqref{eq:vbound}, \eqref{eq:TTC}, \eqref{eq:forwardforcebound}, \eqref{eq:forcebound} \text{ and } \eqref{eq:terminalv}
\end{align}
\end{subequations}	
\end{problem}   
where \eqref{eq:TTC2} and \eqref{eq:opposite} are not imposed. 

Solving OCP \ref{prob:ocp1} yields a set of optimal (non-conservative) CAV trajectories from which the IC can determine the crossing order for the lower-level coordinator. The main idea is motivated, as see in \eqref{eq:TTC2} and \eqref{eq:opposite}, by the fact that vehicles with intersected trajectories within the MZ should follow a crossing order based on their optimized entry times at the MZ. Otherwise, for those vehicles with no intersection between their trajectories in the MZ, 
their orders are determined by their MZ exit times. As such, the implementation steps after solving \cref{prob:ocp1} are performed as follows.

\begin{enumerate}[leftmargin=*,label=Step \arabic*:]
\item Sort CAVs based on their optimized entry times at the MZ, and let $\mathcal{N}_i$ denote the resulting order and denote with $(\mathcal{N}_i)^k$ the k-th element in the order.
\item Label successive CAVs, i.e. $(\mathcal{N}_i)^k$ and $(\mathcal{N}_i)^{k+1}, \,\forall k = 0,\cdots, N\!-\!1$, with ``with collision potentials'' or ``no collision potentials'', depending on their entry positions and intentions.
\item Sort CAVs based on their MZ exit times, and let $\mathcal{N}_o$ denote the resulting order.
\item Swap the orders of $(\mathcal{N}_i)^k$ and $(\mathcal{N}_i)^{k+1}$, $\forall k \!=\! 0,\cdots, N\!-\!1$ in $\mathcal{N}_i$, if they have reversed orders in $\mathcal{N}_o$ ($(\mathcal{N}_i)^{k+1}$ leaves the MZ ahead of $(\mathcal{N}_i)^k$) and they are labelled ``no collision potentials''.
\item By repeating Step 4 for all successive CAV pairs in $\mathcal{N}_i$ with regard to $\mathcal{N}_o$, the crossing order $\mathcal{N}$ is obtained.
\end{enumerate}

\subsection{Trajectory Optimizer} 
Given the intersection crossing order obtained from \cref{prob:ocp1}, the trajectory of each CAV can be optimized by solving \cref{prob:ocp} but with \eqref{eq:TTC2}, \eqref{eq:opposite} enforced based on the new agreed order $\mathcal{N}$.
\begin{problem}\label{prob:ocp2}
	\begin{subequations}\label{eq:OCP2}
\begin{align}
   %\text{minimize:}\quad
    & \hspace{-5mm}\mathop {\min}\limits_{\mathbf{u}} \hspace{4mm} J(\mathbf{x},\mathbf{u})
     \label{eq:J2_ocp2} \\
    \textbf{s.t.}:\quad
    & \eqref{eq:dynamic_dt}, \eqref{eq:syst}, \eqref{eq:vbound}, \eqref{eq:TTC}, \eqref{eq:TTC2}, \eqref{eq:opposite}, \eqref{eq:forwardforcebound},\eqref{eq:forcebound} \text{ and } \eqref{eq:terminalv}
\end{align}
\end{subequations}	
\end{problem} 

It is noteworthy that \cref{prob:ocp1} and \cref{prob:ocp2} are difficult to tackle due to the presence of the nonconvex electric energy consumption model \eqref{eq:battery_usage}, and the nonlinear vehicle longitudinal model \eqref{eq:dynamic_dt} and \eqref{eq:syst}. In this regard, the convex reformulation of \cref{prob:ocp1} and \cref{prob:ocp2} is discussed in the next Section.

\begin{remark}
Given a crossing order obtained in the upper level, there might be a case where no feasible solution can be found in the lower level due to the discrepancy between the upper and lower optimization problems. This can be addressed by recursively solving the lower-level problem with continuously reducing $v_{\min}$, which then terminates when a valid solution is found.  
\end{remark}

%----------------------------------------------------------------------------------------
\section{Convex Problem Formulation and Benchmark Solutions} \label{sec:convex}
In this Section, a convex approximation that yields a feasible upper-bound solution of \eqref{eq:OCP} is introduced, and then a lower-bound solution and a conventional baseline solution (obtained from the simple lossless model in the literature) are also defined for benchmarking purposes. 
Some conceptual preliminaries are introduced before the convex reformulation. As the differences of \cref{prob:ocp1} and \cref{prob:ocp2} only consist in two linear constraints \eqref{eq:TTC2}, \eqref{eq:opposite} that are independent of the convexification, we will use \cref{prob:ocp1} as a representative example for the analysis; the same analysis also applies for \cref{prob:ocp2}.

An SOCP is a convex optimization problem of the form:
\begin{equation}\label{eq:standardSOCP}
\begin{aligned}
     \min\limits_{\mathbf{z}}&\,\bm{c}^T\mathbf{z}\\
     \textbf{s.t.}:&\,\bm{F}\mathbf{z}=\bm{h}\\
     &\,||\bm{A}_{i}\mathbf{z}+\bm{r}_i||_2\leq
     \bm{\alpha}_i^T\mathbf{z}+\bm{\beta}_i,\,i=1,\ldots
\end{aligned}
\end{equation}
where $\mathbf{z} \in \mathbb{R}^n$ is the optimization variable and ${\bm c}$, $\bm{F}$, $\bm{h}$, ${\bm A}_i$, $\bm{r}_i$, $\bm{\alpha}_i$, $\bm{\beta}_i$ are problem parameters. $||\cdot||_2$ is the standard Euclidean norm, and the associated constraint, 
\[
||\bm{A}_{i}\mathbf{z}+\bm{r}_i||_2\leq
     \bm{\alpha}_i^T\mathbf{z}+\bm{\beta}_i \,,
\]
is called a second-order cone constraint.

\subsection{Proposed solution (SOCP-UB)}\label{sec:socp}
The SOCP formulation of \cref{prob:ocp1} is carried out in three steps: 1) reformulation of the objective function, 2) state transformation to linearize the longitudinal dynamics of each CAV, and 3) reformulation of the nonconvex state constants resulting from 2). Certain approximations are applied in 1)-3) under the condition that the approximated problem is feasible to the original problem.

\begin{figure}[htb!]
\centering
\subfigure{
\centering
\includegraphics[width=.8\columnwidth]{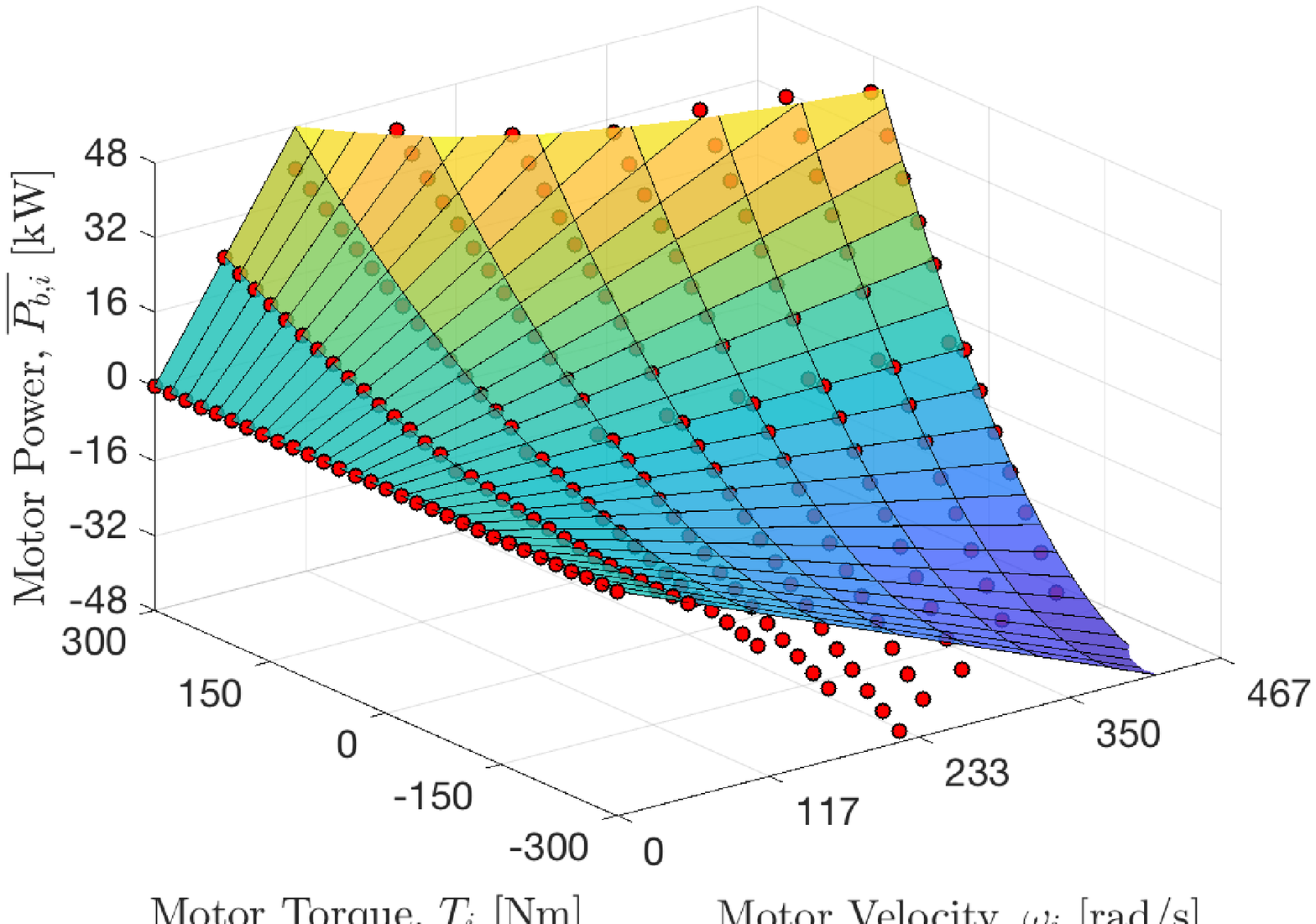} 
}\\
[-2.5ex]
\subfigure{
\centering
\includegraphics[width=.8\columnwidth]{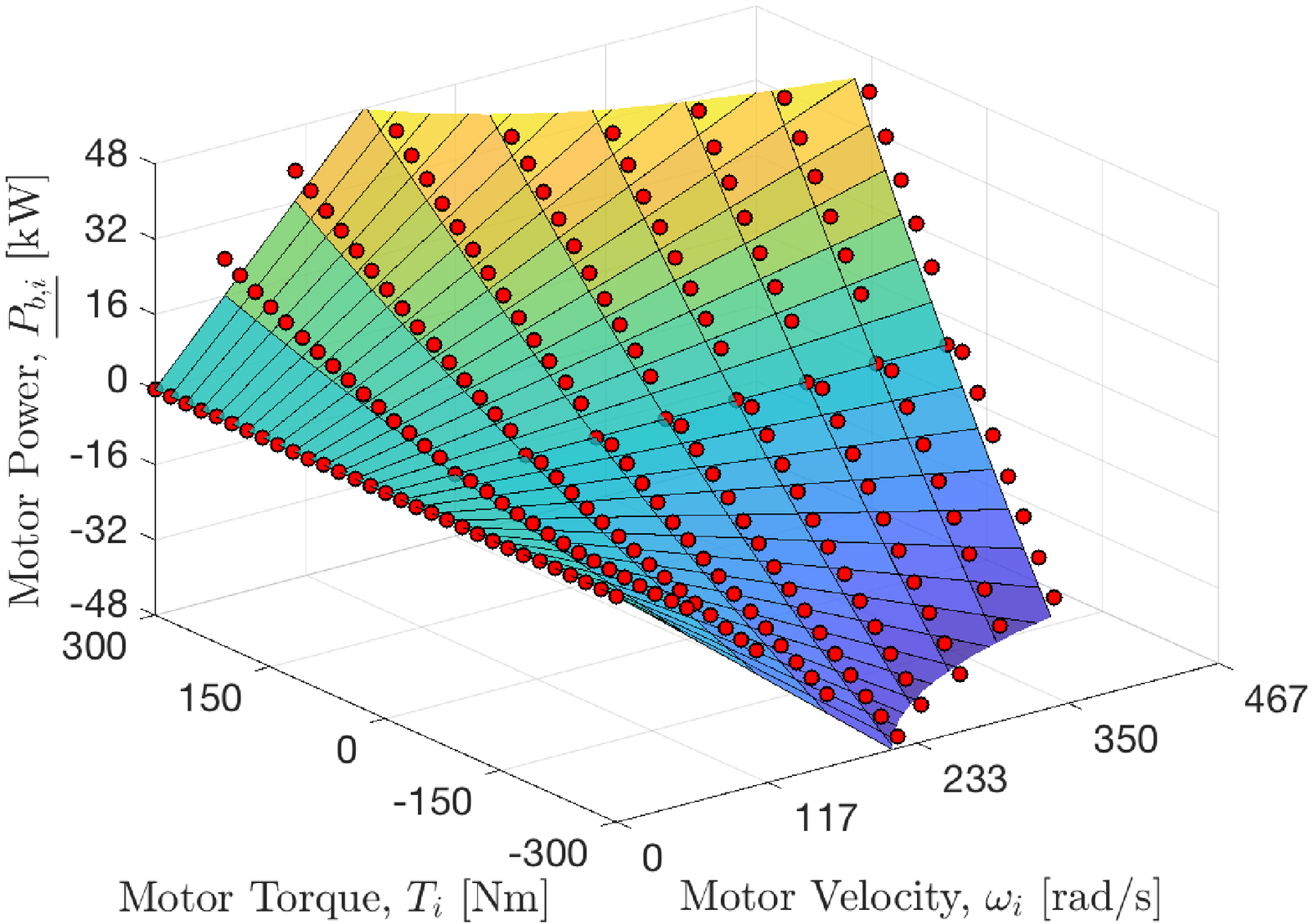} 
}\\[-2ex]
\caption{Nonlinear regression of the motor power data (red dots, calculated based on the efficiency map shown in Fig.~\ref{fig:motormap} using \eqref{eq:Pm2}) by using upper (Top) and lower tangential fitting (Bottom) for SOCP-UB and SOCP-LB, respectively. The R-square fit for the upper case is 94.33\% and 95.13\% for the lower case, respectively.}
\label{fig:Pbfitting}
\end{figure}

%------------------------------------------------------------------------------------------
\paragraph{Step 1}
After the substitution in \eqref{eq:battery_usage}, the electric power defined in \eqref{eq:Pm2} becomes a nonconvex model for battery energy usage due to the presence of $F_{t,i}^2(s)/v_i(s)$. This paper proposes an immediate solution to convexify the resulting battery energy model by replacing the power model \eqref{eq:Pm2} with: 
\begin{align}\label{eq:Pbapprox_upper} 
\overline{P}_{b,i}&=\overline{b}_1F_{t,i}^2v_i+\overline{b}_2F_{t,i}v_i+\overline{b}_3v_i,
%\\
%\underline{P}_{b,i}&=\underline{b}_1F_{t,i}(s)^2v_i+\underline{b}_2F_{t,i}v_i+\underline{b}_3v_i,\label{eq:Pbapprox_lower} 
\end{align}
that can be made to fit tangentially from above (as shown in the top plot in Fig. \ref{fig:Pbfitting}) the results of the battery power according to \eqref{eq:Pm2} calculated based on the efficiency map in Fig.~\ref{fig:motormap}. The fitting parameters $\overline{b}_1$, $\overline{b}_2$, $\overline{b}_3$ are obtained by solving the following constrained optimization problem: 
\begin{subequations}\label{eq:fittedpower}
\begin{align}
& \min\limits_{\overline{b}_1\overline{b}_2\overline{b}_3}\,\left\lVert \overline{P}_{b,i}(v_i,F_{t,i})-P_{b,i}(v_i,F_{t,i})\right\rVert_2 \label{eq:linearapprox1xx}\\
\textbf{s.t.:}\quad 
&\overline{P}_{b,i}(v_i,F_{t,i})-P_{b,i}(v_i,F_{t,i})\geq0 \,.  \label{eq:linear4xx}
\end{align}
\end{subequations} 
which ensures $\overline{P}_{b,i}(v_i,F_{t,i})$ is an upper bound to the optimal power of the original problem \cref{prob:ocp1}. By substituting \eqref{eq:Pbapprox_upper} into \eqref{eq:battery_usage}, the battery energy usage integral of a single CAV for the SOCP-UB problem can be rewritten in a convex quadratic form:
\begin{equation}\label{eq:battery_usageapprox}
\overline{J}_{b,i} =\int_{0}^{2L+\delta(d_i)} \overline{b}_1F_{t,i}(s)^2+\overline{b}_2F_{t,i}(s)+\overline{b}_3\,ds.
\end{equation}

\paragraph{Step 2}
To linearize the nonlinear dynamics \eqref{eq:dynamic_dt} and \eqref{eq:syst}, a change of variable is performed, $v_i \rightarrow E_i$, where $E_i(s)$ is the kinetic energy of CAV $i$ defined by $E_i(s)=\frac{1}{2} m v_i^2(s)$.
As such, \eqref{eq:dynamic_dt} and \eqref{eq:syst} can be rewritten as:
\begin{subequations}\label{eq:ocp-s2-1}
\begin{align}
    &    \frac{d}{d s}E_i(s) =   F_{t,i}(s)+F_{b,i}(s) - F_r - 2\,\frac{f_d}{m}\,E_i(s)\,, \label{eq:sysE}  \\
    & \frac{d}{d s}t_i(s) =\displaystyle  \frac{1}{\sqrt{2E_i(s)/m}},\quad i\in \mathcal{N}, \label{eq:syst2}
\end{align}
\end{subequations}
where \eqref{eq:sysE} becomes linear with respect to the transformed state $E_i(s)$ whereas the dynamics of $t_i(s)$ remain nonlinear. Due to the convexity of \eqref{eq:syst2}, it is reasonable to relax the dynamic equation of $t_i(s)$ into a linear differential equation along with a convex inequality constraint and a penalized cost function of \eqref{eq:J}, described as follows:
\begin{subequations}\label{eq:linearapprox}
\begin{align}
   & \frac{d}{d s}t_i(s) = \zeta_i(s)\,,\label{eq:syst4}\\
   & \zeta_i(s) \geq  \frac{1}{\sqrt{2E_i(s)/m}}, \label{eq:syst3}\\
   & \mathop {\min}\limits_{\mathbf{{u}},\bm{\zeta}} \hspace{2mm} J(\mathbf{{x}},\mathbf{{u}},\bm{\zeta}) =\sum_{i=1}^{N} W_1 \tilde{J}_{t,i} +W_2 \overline{J}_{b,i}, \label{eq:J2} 
\end{align}
\end{subequations}
where $\bm{\zeta} \!=\! [\zeta_{1},\zeta_{2},\ldots,\zeta_{N}]^{\top}$ is a vector of auxiliary control variables $\zeta_i(s)$, and $\tilde{J}_{t,i}$ is the travel time estimated by $\zeta_i$:
\begin{equation}\label{eq:modifiedjt}
\tilde{J}_{t,i} = \int_0^{2L+\delta(d_i)}\!\zeta_i(s) ds\,.
\end{equation}
With $\tilde{J}_{t,i}$ minimized, the control variable $\zeta_i(s)$ intends to find its minimum boundaries (i.e. $\zeta_i(s)\!=\!{1}/{\sqrt{2E_i(s)/m}}$) as solutions. For all the practical scenarios of interest in the present work, it has been found that the present formulation yields a tight solution for \eqref{eq:syst3}, which is considered adequate for the present purposes. Under certain initial conditions, for example, the $i+1$th CAV has proximate arrival time and much higher entering speed at the CZ compared to the $i$th CAV, the tightness of \eqref{eq:syst3} may be violated. Such conditions are unusual in practice and not directly addressed in the present work, however, when they are detected it can be envisaged that control within the ComZ for CAVs $i+1$ and the subsequent CAVs can enable them to arrive at the CZ with valid initial conditions.

To deal with the square root in the non-quadratic constraint \eqref{eq:syst3}, both sides of the constraint are squared yielding,
\begin{equation}\label{eq:zeta_square}
\zeta_i(s)^2 E_i(s) \geq m/2
\end{equation}
which is equivalent to~\cite{alizadeh2003second}:
\begin{subequations}\label{eq:socp1}
\begin{align}
    & E_i(s) \geq \rho_{1,i}^2(s) \label{eq:socp3}\\
   & \zeta_i(s) \geq \rho_{2,i}(s) \label{eq:socp2}\\
    &\rho_{1,i}(s)\rho_{2,i}(s)\geq \sqrt{(m/2) }\label{eq:socp4}\\
    &\rho_{1,i}(s)\geq0,\quad \rho_{2,i}(s)\geq0 \label{eq:socp5}
\end{align}
\end{subequations}
with $\rho_{1,i}(s)$ and $\rho_{2,i}(s)$ auxiliary control variables. 

The kinetic energies $E_i$ are bounded due to the boundedness of the permissible speed limits \eqref{eq:vbound}:  
\begin{equation}
    E_{\min} \leq \,E_i(s)\, \leq E_{\max},\,i\in \mathcal{N},
    \label{eq:Ebound}
\end{equation}
where $E_{\min}\!=\!\frac{1}{2}mv_{\min}^2$ and $E_{\max}\!=\!\frac{1}{2}mv_{\max}(s)^2$ are respectively determined by the velocity limits $v_{\min}$ and $v_{\max}$. 
 
\begin{figure}[htb!]
\centering
\includegraphics[width=.95\columnwidth]{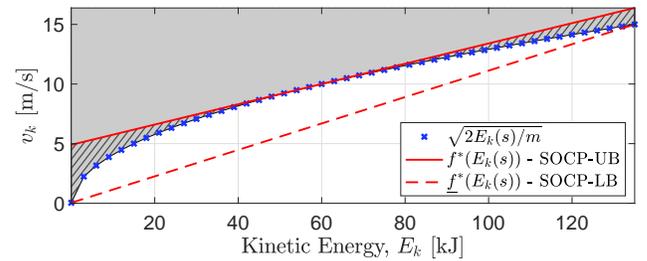}\\[-2ex]
\caption{The solid line shows the linearly approximated relationship between kinetic energy and velocity. The grey region denotes the feasible set and the shaded grey areas indicate the sacrificed feasibility due to the artificial conservativeness. The dashed line shows a nonconservative envelope approximation, which allows to generate a lower bound solution for benchmarking purposes.}
\label{fig:eq8fitting}
\end{figure}

%-----------------------------------------------------------------
\paragraph{Step 3}
After replacing $v_i(s)$ with $E_i(s)$, the rear-end collision avoidance constraint \eqref{eq:TTC} becomes:
\begin{multline}\label{eq:TTC_nonconvex}
 t_k(s)\!-\!t_i(s+l) \\
 > \max\left(\frac{\sqrt{2E_k(s)/m}}{|a_{\min}|}-\frac{\sqrt{2E_i(s+l)/m}}{|a_{\min}|}, t_{\delta} \right),
\end{multline}
which is equivalent to 
\begin{subequations}
\label{eq:TTC_nonconvex12}
\begin{align}
 &t_k(s)\!-\!t_i(s+l) 
 >t_{\sigma} \label{eq:TTC_nonconvex1}\\
& t_k(s)\!-\!t_i(s+l) 
   >\frac{\sqrt{2E_k(s)/m}}{|a_{\min}|}-\frac{\sqrt{2E_i(s+l)/m}}{|a_{\min}|} \label{eq:TTC_nonconvex2}
 \end{align}
\end{subequations}
As it can be noticed, \eqref{eq:TTC_nonconvex2} yields a noncovex feasible set. An immediate and effective solution to convexify the  feasible region is to linearize the nonlinearity induced by the term $\sqrt{2E_k(s)/m}$ (see Fig.~\ref{fig:eq8fitting}). Let us consider $f(E_k(s))$ a linear approximation of the velocity $v_k(s) = \sqrt{2E_k(s)/m}$, such that,
\begin{equation}\label{eq:linearapprox2}
  f(E_k(s)) = a_{0}+a_{1}E_k(s)\,,\quad \forall E_k(s)\in [E_{\min},E_{\max}],
\end{equation}
where $a_{0}$ and $a_{1}$ are obtained through a constrained least-squares optimization for $E_k \in [E_{\min},E_{\max}]$:
\begin{subequations}\label{eq:linear}
\begin{align}
& \min\limits_{a_0,a_1}\,\left\lVert f(E_k)-\sqrt{2E_k/m}\right\rVert_2 \label{eq:linearapprox1}\\
\textbf{s.t.:}\quad 
%&f(E_k)> 0 \,,   \label{eq:linear2}\\
&f(E_k)-\sqrt{2E_k/m}\geq0 \,.   \label{eq:linear4}
\end{align}
\end{subequations} 
which is formed to maximize feasibility while preserving convexity. The problem \eqref{eq:linear} is a simple linear regression problem that can be solved effortlessly. It is noteworthy that there exists a unique solution $f^*(E_k(s)) = a_{0}^*+a_{1}^*E_k(s)$ that is tangential to $\sqrt{2E_k(s)/m}$. As such, the feasibility is confined to a convex set with the boundary $f^*(E_k(s))$ rather than $\sqrt{2E_k(s)/m}$, as shown in Fig.~\ref{fig:eq8fitting}.

Considering the linearization performed by \eqref{eq:linear}, the constraint \eqref{eq:TTC_nonconvex2} can be converted to a convex inequality, as follows:
\begin{multline}\label{eq:TTCapprox}
 t_k(s)-t_i(s+l) 
 \!>\! \frac{a_{0}^*\!+\!a_{1}^*E_k(s)\!-\!\sqrt{2E_i(s+l)/m}}{|a_{\min}|}
\end{multline}
which is a relaxed and conservative (see the shaded grey areas in Fig.~\ref{fig:eq8fitting}) constraint with a larger tolerance on the car-following safety distance, particularly during the low and high speed ranges. Note that the nonlinear term of $E_i$ is retained as the linearization of $E_k$ is sufficient to ensure the convexity of the problem. As such, it avoids extra conservativeness that would be introduced if it was also approximated. To preserve more feasibility, one may use a successive convexification algorithm, which is more computationally demanding due to a recursive project-and-linearize procedure \cite{Mao:ifac2017}. Despite using a simple linearization, the proposed algorithm is shown to yield close-to-optimal results as will be verified by a non-conservative lower bounding solution defined in Section~\ref{sec:socpLB&UB}.

Next, it is shown that the non-quadratic constraint \eqref{eq:TTCapprox} can be quadratically reformulated. By rearranging \eqref{eq:TTCapprox}, it holds that:

\begin{equation}\label{eq:TTC_square22}
     \gamma_i(s) > -\frac{\sqrt{2E_i(s+l)/m}}{|a_{\min}|},
\end{equation}
where,
\begin{equation}\label{eq:gammaforshort}
\gamma_i(s) = t_k(s)\!-\!t_i(s+l) \! -\!\frac{a_{0}^*\!+\!a_{1}^*E_k(s)}{|a_{\min}|},  
\end{equation}
is a linear combination of state variables.
If $\gamma_i(s) \!\geq\! 0$ (which indicates the time gap between two consecutive vehicles is sufficiently large for the following vehicle $k$ to fully stop without collision), the inequality \eqref{eq:TTC_square22} holds naturally as $-\sqrt{2E_i(s+l)/m}/|a_{\min}|$ is always negative. On the other hand, if $\gamma_i(s)<0$, it yields, 
\[
      \frac{\sqrt{2E_i(s+l)/m}}{|a_{\min}|} > -\gamma_i(s) (>0),
\]
which, by squaring both sides, is equivalent to, 
\[
      \frac{2E_i(s+l)/m}{a_{\min}^2} > \gamma_i(s)^2.
\]
After rearrangement, it yields:
\begin{equation}\label{eq:TTC_square5}
      2E_i(s+l) > m a_{\min}^2 \gamma_i(s)^2,
\end{equation}
which is the quadratic counterpart of \eqref{eq:TTCapprox}.

%***************************************
Thus, \cref{prob:ocp1} can be transformed into the following optimization problem:
\begin{subequations}\label{eq:convexformulation}
\begin{align}
 &\min\limits_{\tilde{\mathbf{u}}}\,\tilde{J}=\sum_{i=1}^{N} W_1 \tilde{J}_{t,i} + W_2 \overline{J}_{b,i} \label{eq:convexformulation1}\\
\textbf{s.t.:}\quad 
 \begin{split}
  \frac{d}{ds}\mathbf{x}(s) = f(\mathbf{x},\,\mathbf{\tilde{u}},s)\,, \\
\end{split} \label{eq:convexformulation2}\\
&{\bm \psi}(\mathbf{x}(s),\mathbf{\tilde{u}}(s))\leq 0 \,,  \label{eq:convexformulation3}\\
&{\bm \phi}(\mathbf{x}(0),\mathbf{x}(2L+\delta(d_i)))= 0 \,,   \label{eq:convexformulation4}
\end{align}
\end{subequations} 
where $\tilde{\mathbf{u}}{\triangleq} [\mathbf{u},\bm{\zeta},\bm{\rho}]^{\top}$ and $\bm{\rho} {\triangleq} [\rho_{1,1},\ldots,\rho_{1,N},\rho_{2,1},\ldots,\rho_{2,N}]^{\top}$ and $\tilde{J}$ is expressed in quadratic form with $\tilde{J}_{b,i}$ and $\tilde{J}_{t,i}$ defined in \eqref{eq:battery_usageapprox} and \eqref{eq:modifiedjt}, respectively. The dynamic constraints \eqref{eq:convexformulation2} are formed by \eqref{eq:sysE} and \eqref{eq:syst4}. The inequality constraints \eqref{eq:convexformulation3} include the state constraints \eqref{eq:Ebound}, \eqref{eq:TTC_nonconvex1}, \eqref{eq:TTC_square5} (whereas the reformulated \cref{prob:ocp2} also involves the two linear constraints \eqref{eq:TTC2}, \eqref{eq:opposite}) and the control limits \eqref{eq:forwardforcebound}-\eqref{eq:forcebound}, \eqref{eq:socp1}. Finally, the problem is completed by the boundary condition:
\begin{equation*}\label{eq:terminalE}
  E_i(2L+\delta(d_i)) = \frac{1}{2}m\bar{v}^2 ,\,\,\forall i\in \mathcal{N}
\end{equation*}
that is inferred from \eqref{eq:terminalv}. Since the objective function and  constraints are quadratic, this problem can be immediately turned into an SOCP \cite{alizadeh2003second}.

\subsection{Benchmark solutions}\label{sec:socpLB&UB}
To provide an indication of and therefore evaluate how far a solution is from the true optimal, it is important to develop a lower bounding formulation of OCP \eqref{eq:OCP}. In addition, a baseline solution with the FIFO policy is also solved for further comparison. 

\paragraph{Lower bounding SOCP (SOCP-LB)}
\label{par:SOCP-LB}
Note that the solution of \cref{prob:ocp1} yields a lower bound on the objective function value compared to the global optimum of \cref{prob:ocp}. 
It is proper to consider SOCP-LB as the solution of \cref{prob:ocp1}. To preserve its non-conservativeness, convexification of \cref{prob:ocp1} in such a case requires tight approximations from below of the root term ${\sqrt{2E_k(s)/m}}$ in \eqref{eq:TTC_nonconvex} and of the battery power relationship in \eqref{eq:Pm2}. An immediate solution for the former approximations to connect the two end points ${\sqrt{2E_{\min}/m}}$ and  ${\sqrt{2E_{\max}/m}}$ of the trajectory of ${\sqrt{2E_k(s)/m}},\forall E_k \in [E_{\min},E_{\max}]$, yielding a straight line $\underline{f}^*(E_k(s))$ as shown in Fig~\ref{fig:eq8fitting}. In this context, the problem is solved over the entire feasible set plus a small portion of infeasible set whereby the solution (possibly infeasible) guarantees the same or smaller objective function values compared to the true optimal of the OCP~\eqref{eq:OCP}.

For the latter approximation, a lower tangential fitting to the battery power is performed (see Fig.~\ref{fig:Pbfitting}) to obtain:
\[
\underline{P}_{b,i}=\underline{b}_1F_{t,i}(s)^2v_i+\underline{b}_2F_{t,i}v_i+\underline{b}_3v_i,
\]
where $\underline{b}_1$, $\underline{b}_2$, $\underline{b}_3$ can be obtained analogously to \eqref{eq:fittedpower}.

Following the same steps in the SOCP-UB introduced in Section~\ref{sec:socp}, it is straightforward to formulate SOCP-LB simply by replacing $a_0^*$ and $a_1^*$ in \eqref{eq:TTCapprox} with the coefficients of the straight line $\underline{f}^*(E_k(s))$, and by substituting $\overline{P}_{b,i}(s)$ in \eqref{eq:battery_usageapprox} with $\underline{P}_{b,i}(s)$, as follows:
\begin{equation}
\underline{J}_{b,i} =  \int_{0}^{2L+\delta(d_i)} \underline{b}_1F_{t,i}(s)^2(s)+\underline{b}_2F_{t,i}(s)+\underline{b}_3(s) \,ds.
\end{equation}

\paragraph{Baseline solution (SOCP-Baseline)}
The baseline solution consist in solving \cref{prob:ocp2} by following the SOCP framework proposed in Section~\ref{sec:socp} with the constraints \eqref{eq:TTC2} and \eqref{eq:opposite} set in line with the FIFO protocol, as with the approaches widely used in the literature \cite{hadjigeorgiou19,Malikopoulos:2018}.

\subsection{Energy Consumption Evaluation} \label{subsec:energyconsumevaluation}
For a fair comparison between different methodologies, the battery energy consumption for all methods is evaluated using the nominal motor efficiency map in Fig.~\ref{fig:motormap}. For a single CAV $i$, it follows $E^*_{\text{Bat},i}\!=\!\int_{0}^{2L+\delta(d_i)}F_{\text{Bat},i}^*(s)\,ds$, where
\begin{equation}\label{eq:Jb_evaluation}
\begin{aligned}
F_{\text{Bat},i}^*(s)\!=\! \left\{
 \begin{array}{ll}
   \displaystyle \!\!\!\frac{F_{t,i}^*(s)}{\eta_m(F_{t,i}^*(s),E_{i}^*(s))}, \hspace{2cm} \forall F_{t,i}(s)\geq0, \\
   \displaystyle
    \!\!\!{(F_{t,i}^*(s)-F_{b,i}^*(s))}{\eta_m(F_{t,i}^*(s),E_{i}^*(s))},\\\hspace{4.7cm} \forall F_{t,i}(s)<0.
 \end{array}\right.
 \end{aligned}
\end{equation}
$E_{\text{Bat},i}^*(s)$ is the actual battery energy consumption obtained by following the optimal control actions $F_{t,i}^*(s)$, $F_{b,i}^*(s)$ and the corresponding optimal state $E_{i}^*(s)$ solved in each case, and $\eta_m(F_{t,i}^*(s),v_{i}^*(s))$ is the powertrain efficiency (a look-up table) shown in Fig.~\ref{fig:motormap}.

%-------------------------------------------------------------------------------------------------------------------------------------------------------------------------------------------------------------------
\section{Numerical Results} \label{sec:simulation}
\begin{figure*}%[htb!]
\centering
\includegraphics[width=.98\textwidth]{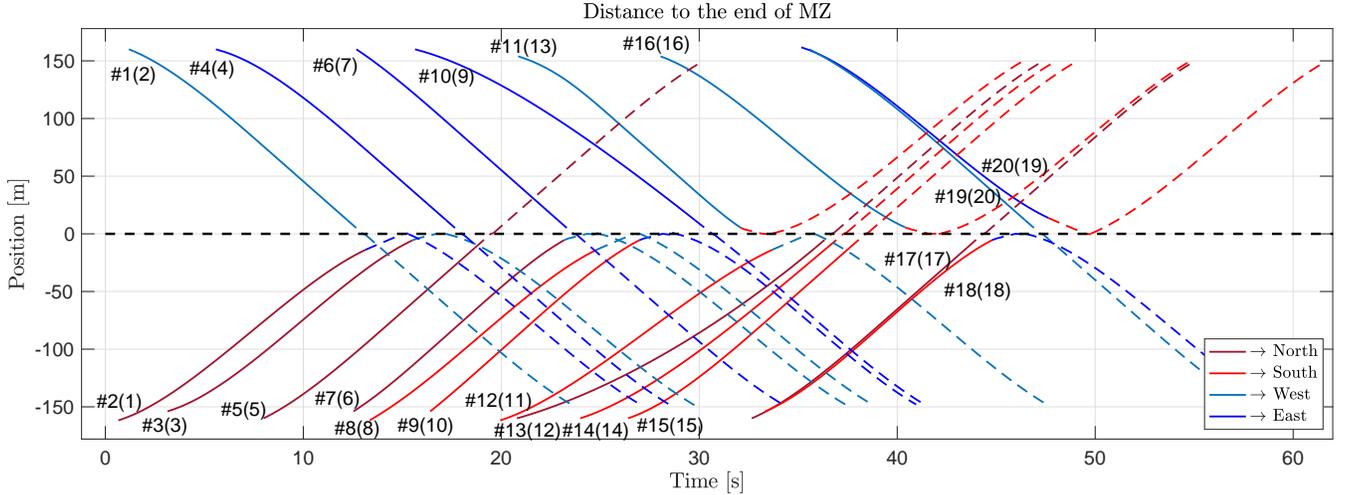}\\[-2ex]
\caption{Trajectories of the first 20 CAVs among 60 CAVs by solving the SOCP-UB with $N\!=\!60$ CAVs at an arrival rate of 750 veh/h per lane. The dashed black line represents the end of the MZ. The dash-colored trajectories correspond to trajectories of all CAVs passing the entry of the MZ. The four vehicle heading directions are denoted with different colors. The numbers with \# denote the crossing order $\mathcal{N}$ and the numbers in the brackets are the arriving order at the CZ. The crossing order for the first 20 CAVs computed by the upper-level scheduler is $\mathcal{N}=\{2,1,3,4,6,5,7,8,10,9,11,12,13,14,15,17,16,19,20,18\}$.
}
\label{fig:distance_example}
\end{figure*}

\begin{figure*}%[htb!]
\centering
\subfigure{
\includegraphics[width=.475\textwidth]{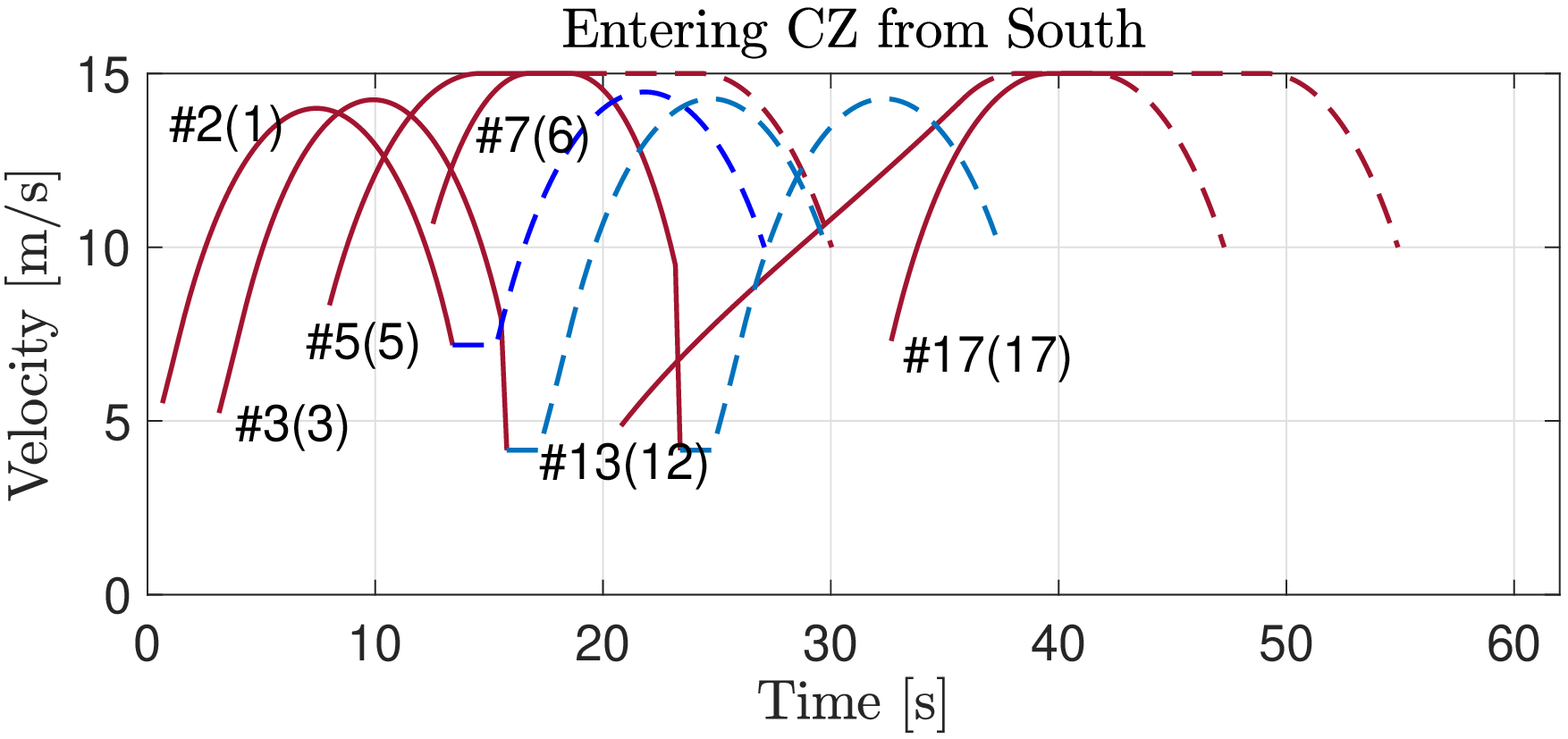}
}
\subfigure{
\includegraphics[width=.475\textwidth]{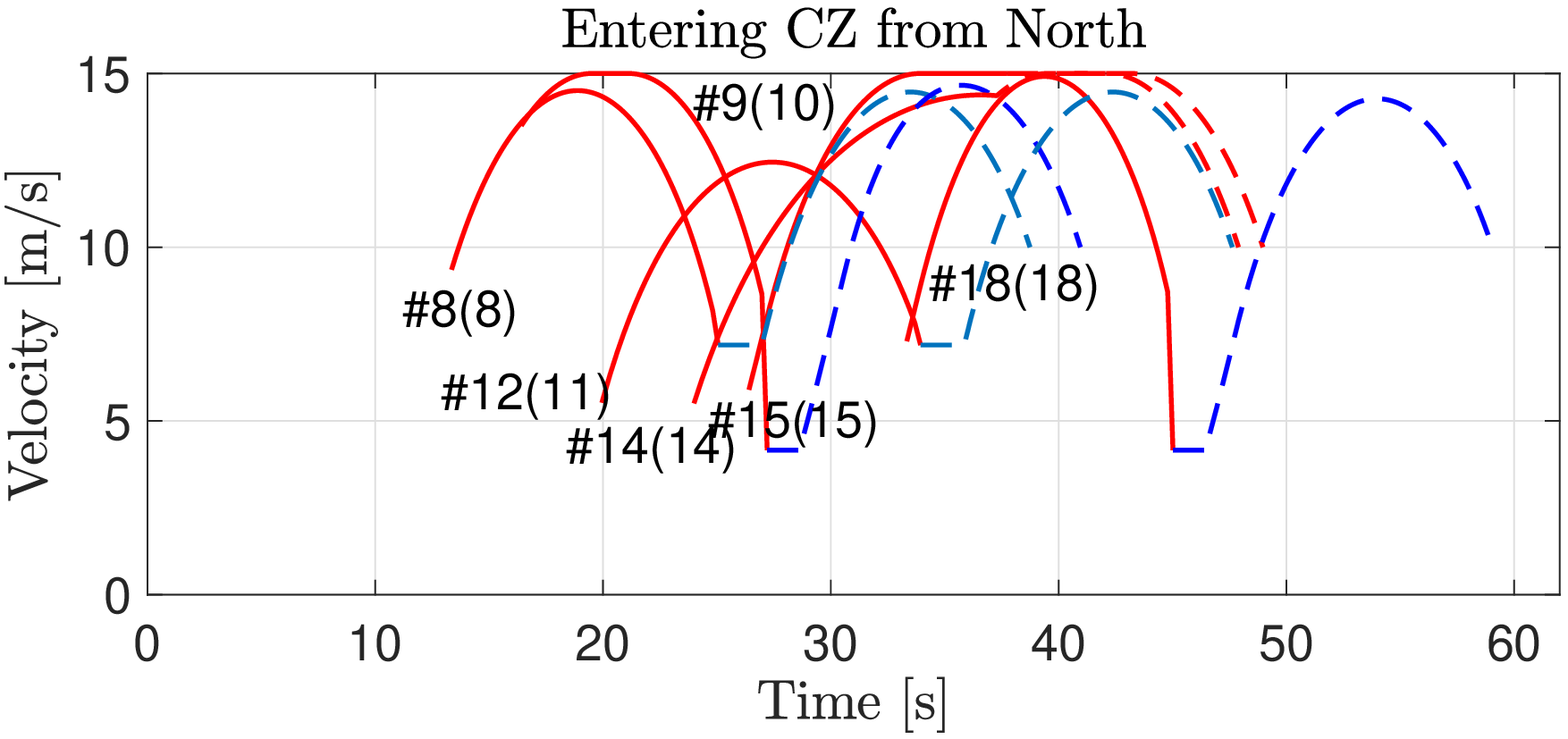}
}\\[-6.5ex]
\subfigure{
\includegraphics[width=.475\textwidth]{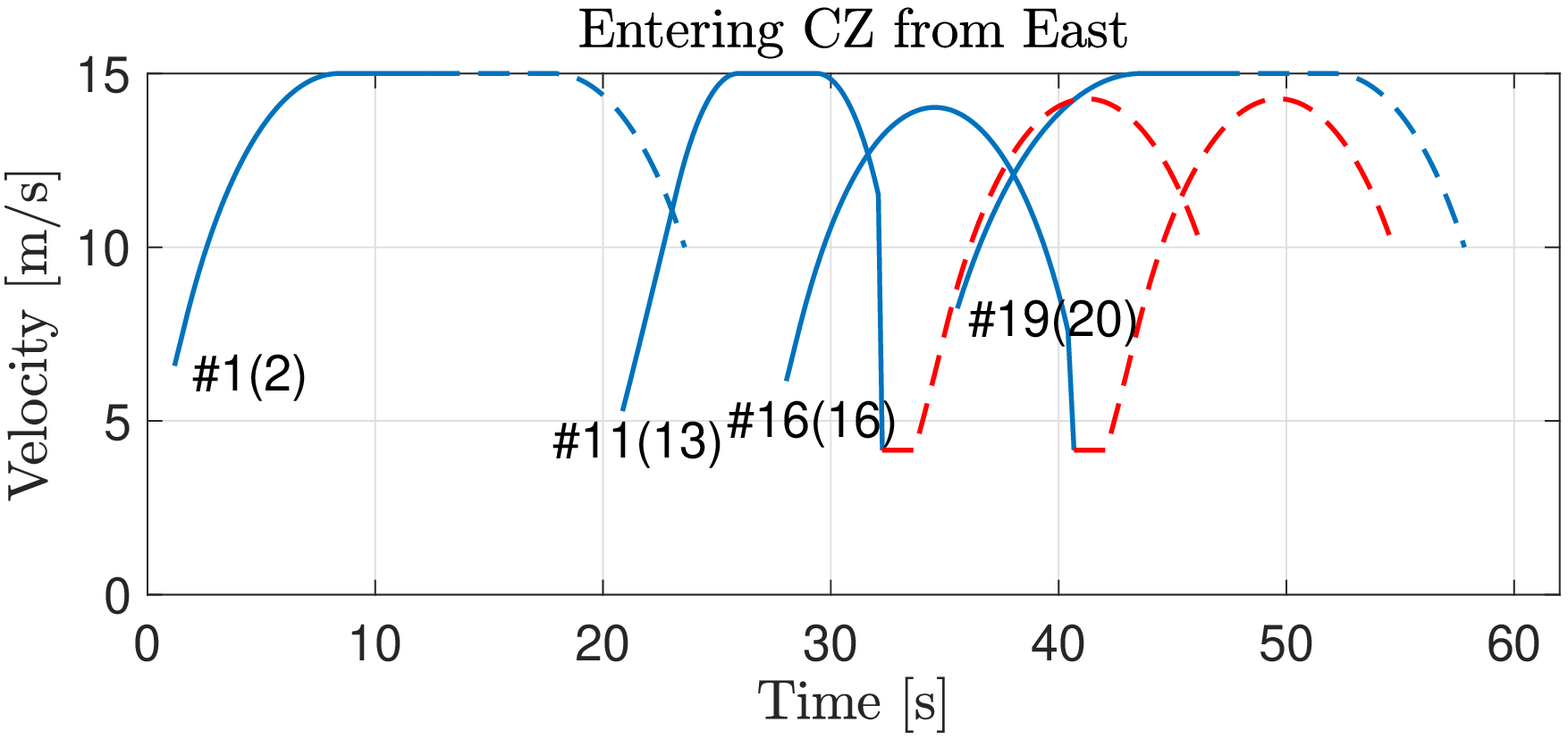}
}
\subfigure{
\includegraphics[width=.475\textwidth]{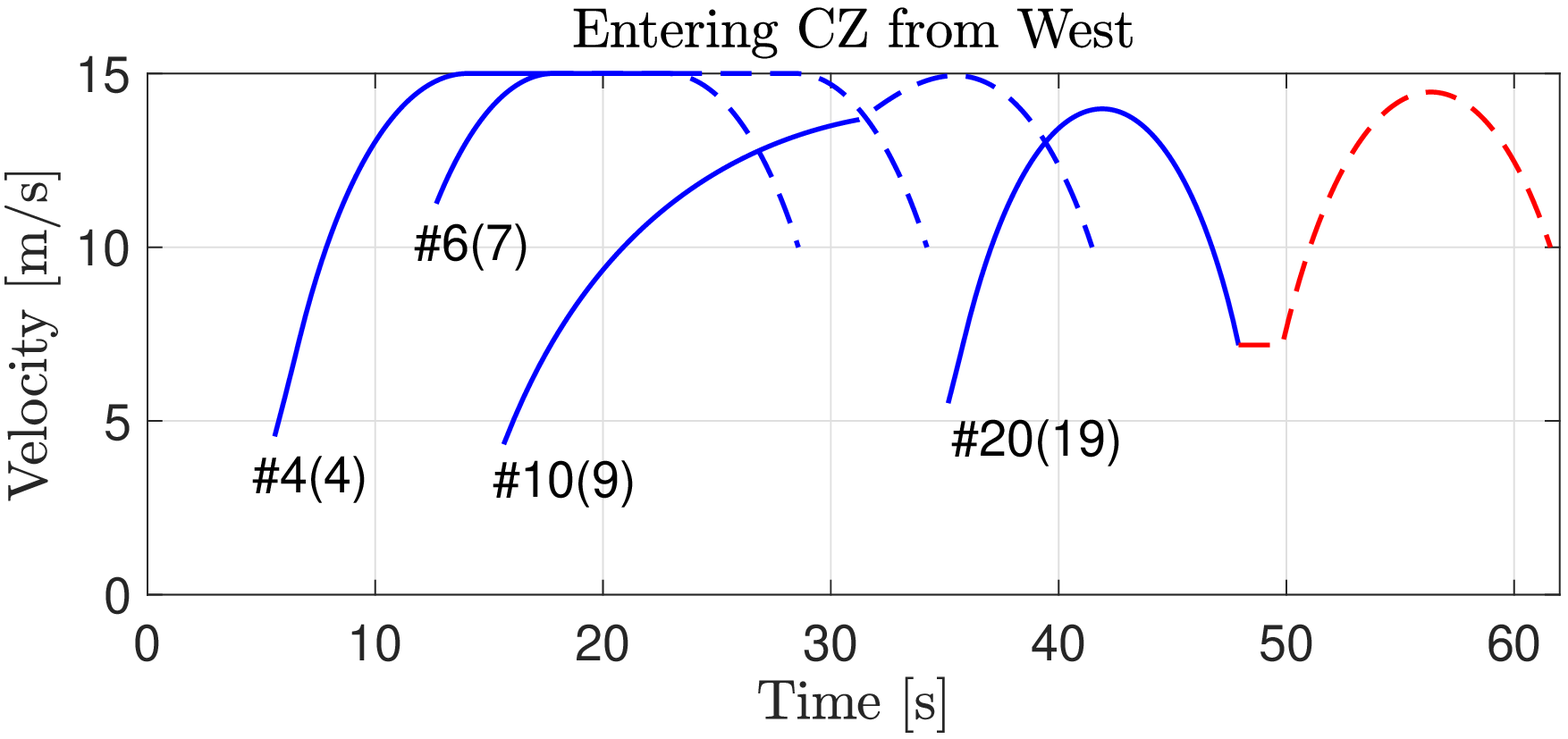}
}\\[-2ex]
\caption{Optimal speed profiles for the first 20 CAVs among 60 CAVs by solving the SOCP-UB at an arrival rate of 750 veh/h per lane with the peak speed limit at $15$~m/s. The solid lines represent the trajectories before vehicles enter the MZ, while the dashed lines show the speed profiles after the vehicle enters the MZ. The numbers with \# denote the crossing order $\mathcal{N}$ and the numbers in the brackets are the arriving order at the CZ.}
\label{fig:velocityprofile}
\end{figure*}

The evaluation of the proposed method is fourfold: 1) the SOCP-UB \eqref{eq:convexformulation} is solved for different weighting combinations $\{W_1,W_2\}$ of \eqref{eq:convexformulation1} under a series of different arrival rates to show the trade-off between energy cost and travel time as well as the impact of the traffic density on the overall optimality; 2) the relationship between the safety margin between the vehicles and energy consumption is examined by solving SOCP-UB for different arrival rates subject to a fixed average travel time; 3) the tightness of the relaxation bounds involved in SOCP-UB is investigated by benchmarking SOCP-UB against SOCP-LB; 4) the performance of the SOCP-UB is compared to the SOCP-Baseline to show the benefit of the proposed coordination scheme over the FIFO policy in terms of optimality.

In the following case studies, we consider an intersection following the layout in Fig.~\ref{fig:Intersection}, with $L\,{=}\,150$\,m and $S\,{=}\,10$\,m. Given the size of the MZ, the turning radii can be calculated as $R_l\!=\!2.50$~m and $R_r\!=\!7.50$~m, and from \eqref{eq:turningspeedlimits},
$v^l_{\max}$ and $v^r_{\max}$ are determined as $v^l_{\max}\!=\!4.16$~m/s and $v^r_{\max}\!=\!7.19$~m/s. $v_{\min}\!=\!0.1$~m/s and $v^f_{\max}\!=\!15$~m/s are also applied. All the vehicles are assumed to leave the intersection at the same terminal speed $\bar{v}\!=\!10\,$m/s. {In \eqref{eq:TTC}, the time safety margin is set to $t_{\delta}\!=\!\Delta s/v_{\max}\!=\!0.13$\,s with $\Delta s\!=\!2$\,m the sampling space, to accommodate the rapid velocity changes within a sampling space interval.}
Without loss of generality, the control problem is initialized with randomized initial conditions $v_i(0)$ and $t_i(0)$ for all CAVs subject to the constraints imposed in Assumptions~\ref{ass:initial}. In particular, CAVs' initial speeds follow a uniform distribution within $[v_{\min},\,v_{\max}]$, while their arrival times, $t_i(0)$, follow a Poisson distribution. Moreover, the entry direction and turning decisions, $d_i$ of each CAV are also randomly generated. The SOCP is solved using the convex solver CVX with MOSEK~\cite{cvx} in Matlab on a personal computer with Intel Core i5 2.9 GHz and 8 GB of RAM. 
Table.~\ref{tab:Computational_Time} shows that the average running time for 20 vehicles is less than 3~s, and it is less than 8~s when the vehicle number is increased to 60.
\begin{table}[!ht]
\centering 
\caption{Average, Minimum \& Maximum Computation Time of $100$ Simulation trials with Randomized Initial Conditions for the Proposed Hierarchical Centralized Coordination Scheme.}
\label{tab:Computational_Time} 
\begin{tabular*}{1\columnwidth}{l @{\extracolsep{\fill}}  c @{\extracolsep{\fill}}  c}
\hline
\hline
 Vehicle number, $N$ & 20 & 60  \\
 \hline
 Average Computational time [s] & {2.71} & {7.52}   \\
 Minimum Computational time [s] & {2.59} & {7.30}  \\
 Maximum Computational time [s] & {2.86} & {7.89} \\
\hline
\hline
\end{tabular*}
\end{table}
To present a general traffic flow scenario, the vehicle number in the following cases is set to $N\!=\!60$.

In the first instance, the SOCP-UB is solved at an arrival rate of $750$~veh/h (vehicles per hour) per lane, which is ordinary for practical intersections. 
The weighting factors are set to emphasize more on the travel time term in the objective function. For illustrative purposes, the traveled distance and velocity profiles of only the first 20 CAVs are shown as the rest of the vehicles exhibit similar patterns of distance and speed profiles.
Fig.~\ref{fig:distance_example} shows the traveled distance profiles. As can be observed, the cooperatively assigned crossing order $\mathcal{N}$ is distinct from the arriving order at the CZ owing to the upper-level scheduling mechanism. Given order $\mathcal{N}$ defined in the upper-level, the lower-level controller schedules the CAVs such that no CAV violates the rear-end and lateral collision constraints, which verifies the validity of the optimal solution. More specifically, if two vehicles have a potential collision inside the MZ, the following one will not be allowed to enter the MZ until the vehicle ahead has left (see \eqref{eq:TTC2}), such as vehicles {\#19} and {\#20}, where the number with \# denotes the vehicle order in $\mathcal{N}$. Conversely, if the paths of two or more vehicles do not intersect, they are allowed to travel inside the MZ at the same time, such as vehicles {\#4} and {\#5}. Moreover, the effectiveness of rear-end collision avoidance can be identified as the solution has no intersections between trajectories of the same color throughout the CZ. 

The optimal speed trajectories of the first 20 vehicles are shown in Fig.~\ref{fig:velocityprofile}, where the profiles are grouped based on the entering direction at the CZ. As it can be seen, the speed trajectory of a CAV highly depends on its decision $d_i$ at the intersection, which can be inferred by the change of color between the solid and dashed parts of the lines depicting the speed profiles. If $d_i\!=\!0$, the path of the vehicle within the CZ is a straight line, and in this context, the CAV accelerates to a cruising speed value and stays at this speed until the exit of the CZ approaches. On the other hand, if a turn is made at the intersection, that is $d_i\!=\!\{-1,1\}$, the optimal speed profiles involve two separate phases, joined by a short period of cruising at the speed limits for cornering inside the MZ (e.g., vehicles {\#1} and {\#2}). In some cases, the speed may not follow the foregoing trajectories due to the compromise on safety requirements. For example, vehicle {\#17} exhibits relatively lower acceleration at the beginning compared to others, and vehicle {\#11} decelerates its speed until it enters the MZ.
Vehicle {\#7} applies additional braking before entering the MZ in order to leave enough space margin to allow the vehicle ahead, {\#6}, to complete its turn. Such a compromise is more noticeable when the arrival times of two consecutive vehicles are close to each other, for example under high traffic density conditions. Moreover, as electric vehicles have recuperative brakes, all CAVs intend to follow a regular profile that involves acceleration to a cruise speed value, followed by a period of constant speed cruising until the exist of CZ (for straight running) or the entry of MZ (if turning is required) approaches, so as to reduce total electric energy consumption \cite{chen:tvt2019}.  A visual demonstration of the optimal solutions can be found {at the link {\blue \href{https://youtu.be/Xmh6pOzlSe0}{https://youtu.be/Xmh6pOzlSe0}}.}

\begin{figure}[htb!]
\centering
\includegraphics[width=\columnwidth]{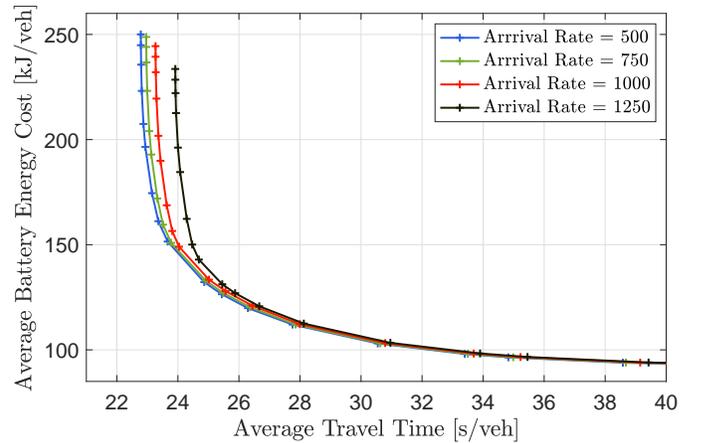}\\[-2ex]
\caption{Trade-off between average battery energy consumption and average travel time for arrival rates from 500~veh/h per lane to 1250~veh/h per lane and for a range of ($W_1,W_2$) pairs.}
\label{fig:tradeoff}
\end{figure}

In order to investigate the impact of traffic density and the trade-off between energy consumption and travel time, the optimal solutions for a series of combinations of the weight factors, $W_1$ and $W_2$, (under the same initial conditions) and for different arrival rates are presented in Fig.~\ref{fig:tradeoff}. 
As it can be seen, the Pareto front results for four different arrival rates indicate that an increase in travel time of approximately {20\%} can lead to an average fuel consumption reduction of {41.7\%}, while further increase in travel time can eventually yield up to {55.6\%} fuel consumption reduction. These results point out the importance of examining the energy-time trade-off, as a small sacrifice in travel time can significantly affect the energy efficiency. The comparison among four arrival rates indicates that the overall optimality deteriorates as the arrival rate increases. The reason is that a higher arrival rate implies a higher traffic density condition, where the motions of vehicles are more restrained by the surrounding vehicles, and therefore, the optimal solution tends to be compromised by collision avoidance requirements. Furthermore, the influence of the arrival rate is more apparent when the average travel time is small. This can be understood that with an emphasis on the travel time minimization, the optimization encourages the CAVs to travel at maximum speed, which yields more restrictive solutions due to the tougher collision avoidance constraints in such cases, and the restrictiveness rises as the arrival rate increases. Finally, it has been found that further decrease in the arrival rate below 500~veh/h makes negligible impact on the optimality, as the traffic is sufficiently sparse to allow free optimization of each velocity trajectory without concession to other vehicles.

To examine the relationship between the safety margin and energy consumption, a comparison of the average time gap and energy consumption is made among cases of different arrival rates while keeping the average travel time fixed. Note that the time gap is defined in \eqref{eq:TTC} and the average is taken only for the subset of vehicles with potential for rear-end collision ($\mathcal{C}_i$ set).
Table~\ref{tab:TTC_performance} presents the results  for the case of a fixed average traveled time of {26.77~s}, which is representative of cases where  arrival rate changes have an influence on energy consumption (see Fig.~\ref{fig:tradeoff}). 
\begin{table}[!ht]
\centering 
\caption{Average Time Gap at Different Arrival Rates with a Fixed Average Traveled Time {26.77~s} with ${t_{\delta}}=\Delta s/v_{\max}=0.13$~s.}
\label{tab:TTC_performance} 
\begin{tabular*}{1\columnwidth}{l @{\extracolsep{\fill}} c@{\extracolsep{\fill}}c@{\extracolsep{\fill}} c @{\extracolsep{\fill}} c}
\hline
\hline
 Arrival Rate [veh/h] & 500 & 750 & 1000 & 1250 \\
 \hline
 Energy Cost [kJ] & {110.93} & {112.97}& {119.95} & {131.34} \\
 Average time gap [s] & {7.14} & {4.90} & {4.11} & {3.74} \\
 Minimum time gap [s] & {0.17} & {0.15} & {0.14} & {0.13 }\\
 Maximum time gap [s] & {20.46} & {14.32} & {13.02} & {12.41}\\
\hline
\hline
\end{tabular*}
\end{table}
As it can be seen, there is an upward trend in the energy cost from {111.43~kJ to 119.95~kJ} as the arrival rate increases from 500~veh/h to 1000~veh/h. Meanwhile, the average time gap decreases from {7.14~s to 4.11~s} for the same arrival rate change. This can be understood that increased traffic density could result in severe congestion and more acceleration/deceleration behavior, and therefore reduced time gap and higher energy consumption. When the arrival rate is 1250~veh/h, the energy consumption is steeply compromised to {131.34~kJ}, as also shown in Fig.~\ref{fig:tradeoff}, which is influenced by the higher number of activation of the limiting time gap (see $t_\delta$ in \eqref{eq:TTC} and also observe that the minimum time gap for this arrival rate in Table~\ref{tab:TTC_performance} reaches  $t_\delta\!=\!0.13$~s).

\begin{figure}[htb!]
\centering
\includegraphics[width=\columnwidth]{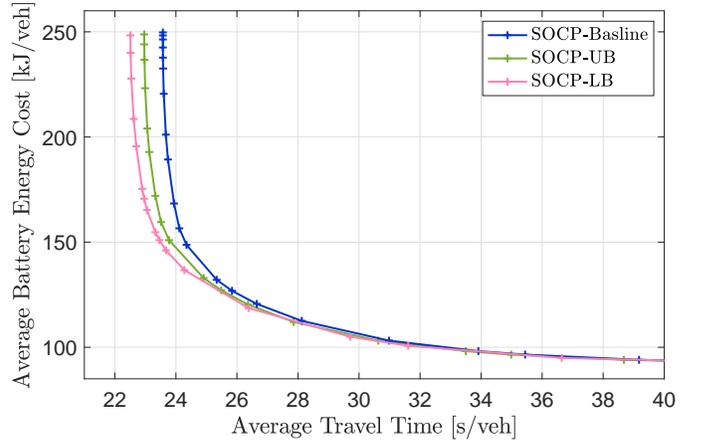}\\[-2ex]
\caption{Comparison of the energy-time cost trade-off between the SOCP-UB and the benchmark solutions at an arrival rate of 750 veh/h per lane.}
\label{fig:optimality_bound}
\end{figure}

The optimality of the SOCP-UB is investigated by comparing its performance with SOCP-LB and SOCP-Baseline introduced in Section~\ref{sec:socpLB&UB}. As shown in Fig.~\ref{fig:optimality_bound}, the solutions of the SOCP-UB are close to the SOCP-LB, which implies the tightness of the linearly approximated bound shown in Fig.~\ref{fig:eq8fitting} and the battery power shown in Fig.~\ref{fig:Pbfitting}. Owing to the increased feasibility in terms of the following distance, the SOCP-LB as compared to the SOCP-UB can reach a more time-efficient solution, which, however, is potentially unsafe (infeasible). As an example, when the average energy cost is {$147$~kJ/veh}, the average travel time is increased by only {1.6\%} for the SOCP-UB as compared to SOCP-LB. 
It can also be observed that when $W_1\gg W_2$, the energy cost becomes negligible in the objective function, thus the optimal speed trajectories derived by the two schemes (SOCP-LB and SOCP-UB) are pushed to the upper speed limits, and the optimality gap is further increased to {2.3\%} at an average energy cost of {$189.5$~kJ/veh}. On the other hand, the Pareto front of the SOCP-UB is always below that of the SOCP-Baseline with a maximum optimality gap at {2.4\%} at the same average energy cost of {$189.5$~kJ/veh}, which highlights the benefits of using a planning-based scheduling method instead of the FIFO policy.  In particular, when the average travel time is {24.74~s}, the SOCP-UB can save up to {21.8\%} energy consumption with respect to the SOCP-Baseline. As the energy cost weight, $W_2$, is gradually increased, the optimality gap among the three schemes becomes negligible. This can be understood by the fact that the CAVs are encouraged to travel at a lower average speed when the emphasis is on energy consumption, resulting in large enough time gaps between CAVs so that their speed trajectories can be freely optimized without being limited by the safety enforcement constraints.

\section{Conclusions}\label{sec:conclusions}
In this paper, the traffic coordination problem at signal-free traffic intersections is addressed for connected and autonomous vehicles. The dynamics of each vehicle are modeled by a realistic longitudinal model in conjunction with an explicitly formulated electric powertrain system, which allows the energy consumption to be accurately estimated. The problem is approached by a hierarchical centralized coordination scheme that aims to minimize a weighted sum of the aggregate electric energy consumption and traveling time required to drive through the junction by sequentially optimizing the passing order and explicit velocity trajectories in two stages. The overall problem is formulated in the space domain, and in this context, the resulting optimal control problems (OCPs) in both stages can be respectively suitably relaxed as convex second-order cone programming (SOCP) problems, which can be solved to optimality efficiently using a standard optimization solver. 

Simulation results verify the validity and computational efficiency of the solution obtained by the proposed control scheme, which enables the method to be implemented using current technologies. To illustrate the trade-off between energy consumption and travel time, a range of cases with different weighting on these two costs are examined and the Pareto front corresponding to different combinations of the two costs is produced. The investigation of the Pareto solutions emphasizes the importance of optimizing their trade-off, as a compromise of {20\%} in travel time could lead to up to {41.7\%} in energy savings. According to the comparison with a valid lower bounding solution of the full original problem, the presented approximation OCP algorithm is able to achieve feasible solutions close to this bounding solution, which further demonstrates the tightness of the convex relaxation employed in the proposed OCP. Finally, the method proposed in this paper is compared to a benchmark solution commonly employed in the literature, obtained using a simple first-in-first-out (FIFO) policy. The proposed technique is found to outperform the benchmark solution with up to an impressive {21.8\%} improvement in terms of energy-saving when travel time in both cases is equalized, and furthermore with the same energy consumption, the method can save up to {2.4\%} travel time.

%----------------------------------------------------------------------
\bibliographystyle{IEEEtran}
\bibliography{paper}

\begin{IEEEbiography}[{\includegraphics[width=1in,height=1.25in,clip,keepaspectratio]{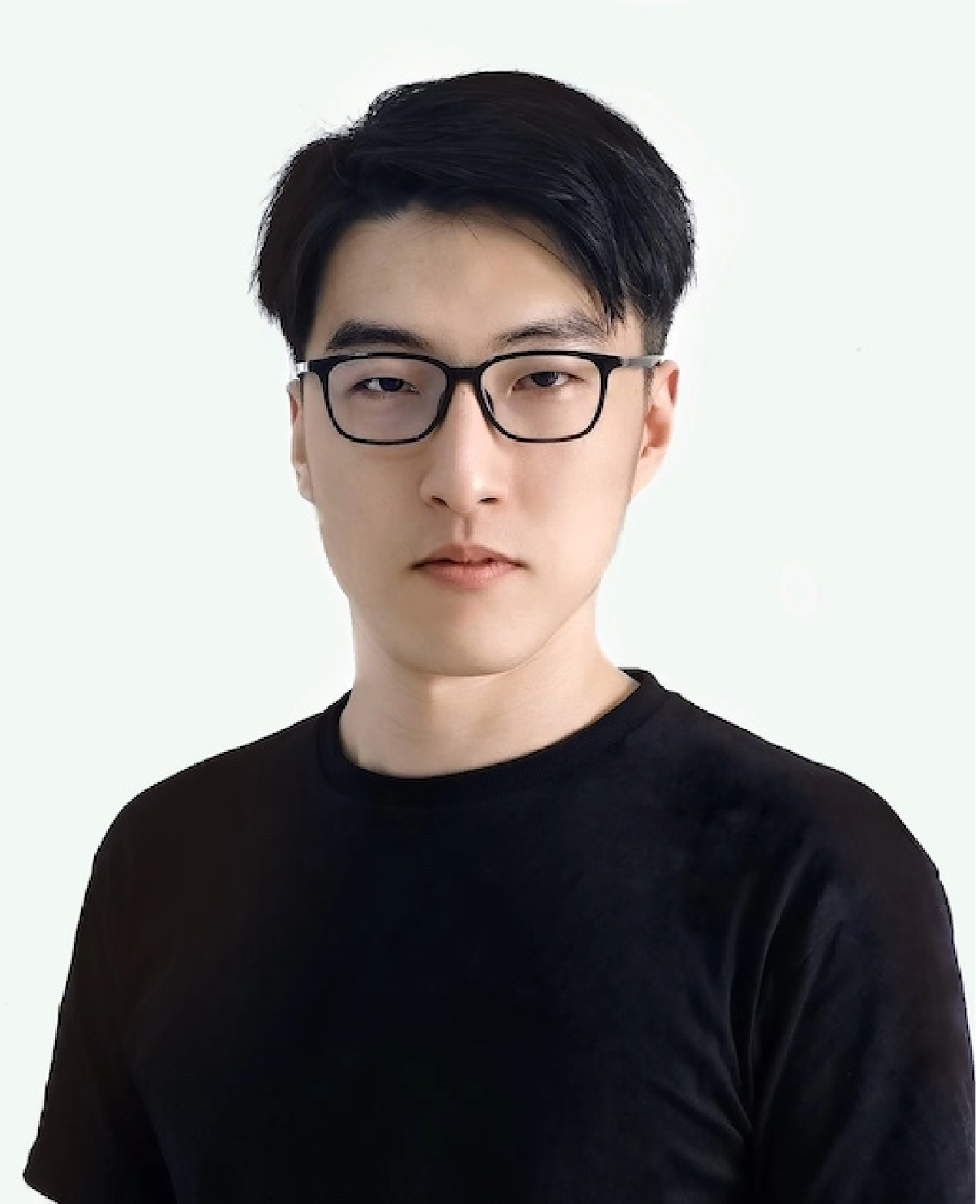}}]
{Xiao Pan} received the B.Sc. degree in Automation in 2017 from Xu Teli School of Beijing Institute of Technology, Beijing, China. In 2018, he received the M.Sc. degree in Control Systems (Distinction) from Imperial College London, London, UK. He is currently working towards his Ph.D. degree with the Department of Electrical and Electronic Engineering, Imperial College London, London, U.K. His research interests include modeling, control, and optimization theory with a specific emphasis on automated and electric vehicles.
\end{IEEEbiography}

\begin{IEEEbiography}[{\includegraphics[width=1in,height=1.25in,clip,keepaspectratio]{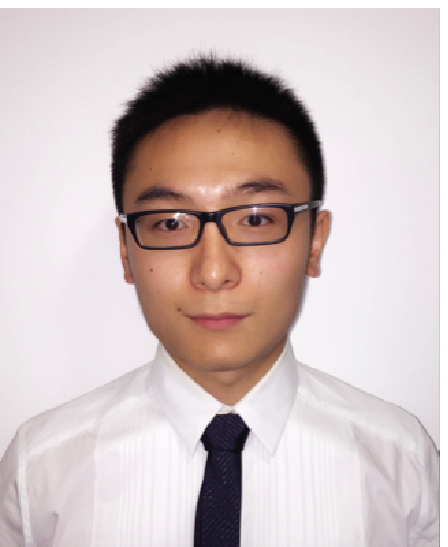}}]
{Boli Chen} (M'16) received the B. Eng. in Electrical and Electronic Engineering in 2010 from Northumbria University, UK. In 2011 and 2015, he respectively received the MSc and the Ph.D. in Control Systems from Imperial College London, UK. Currently, he is a Lecturer in the Department of Electronic and Electrical Engineering, University College London, U.K. His research focuses on control, optimization, estimation and identification of a range of complex dynamical systems, mainly from automotive and power electronics areas.
\end{IEEEbiography}

\begin{IEEEbiography}[{\includegraphics[width=1in,height=1.25in,clip,keepaspectratio]{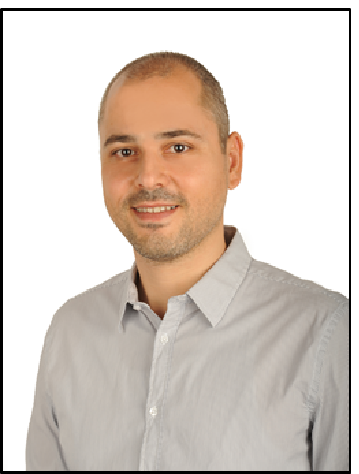}}]
{Stelios Timotheou} is an Assistant Professor at the Department of Electrical and Computer Engineering and a faculty member at the KIOS Research and Innovation Center of Excellence, of the University of Cyprus. He holds a Dipl.-Ing. in Electrical and Computer Engineering (Summa Cum Laude, 2005) from the National Technical University of Athens, an MSc in Communications and Signal Processing (Distinction, 2006) and a PhD in Intelligent Systems and Networks (2010), both from the Department of Electrical and Electronic Engineering of Imperial College London. In previous appointments, he was a Research Associate at KIOS, a Visiting Lecturer at the Department of Electrical and Computer Engineering of the University of Cyprus, and a Postdoctoral Researcher at the Computer Laboratory of the University of Cambridge. His research focuses on monitoring, control and optimization of critical infrastructure systems, with emphasis on intelligent transportation systems and communication systems. He is the recipient of the 2017 Cyprus Young Researcher in Physical Sciences \& Engineering Award, by the Cyprus Research Promotion Foundation.
\end{IEEEbiography}

\begin{IEEEbiography}[{\includegraphics[width=1in,height=1.25in,clip,keepaspectratio]{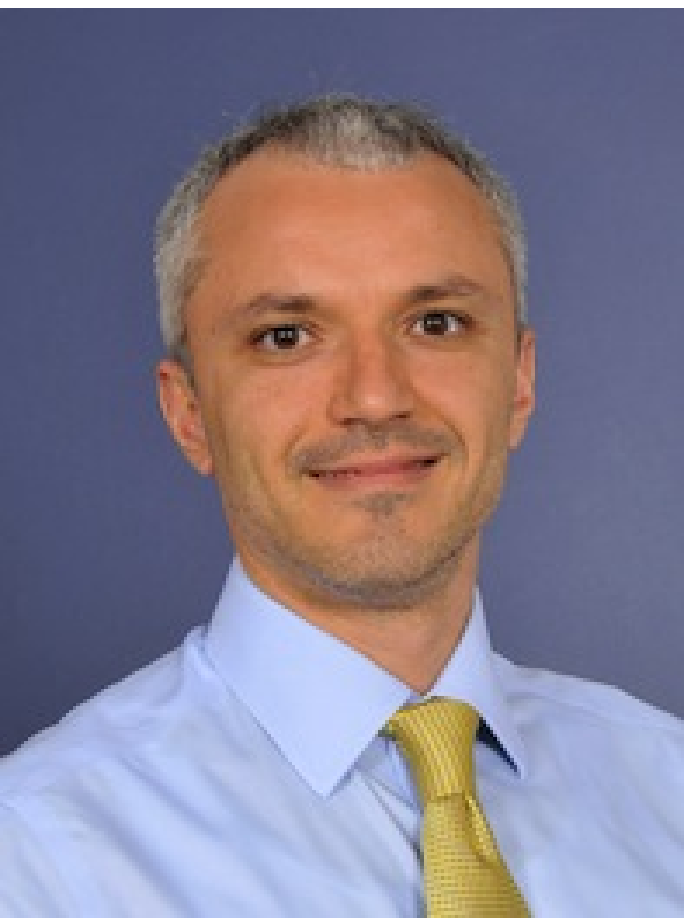}}]
{Simos A. Evangelou} (Senior Member, IEEE) received the B.A./M.Eng. degree in electrical and information sciences from the University of Cambridge, Cambridge, U.K., in 1999, and the Ph.D. degree in control engineering from Imperial College London, London, U.K., in 2004. He is currently a Reader with the Department of Electrical and Electronic Engineering, Imperial College London, London, U.K. He is a Member of IFAC Technical Committee Automotive Control and on the editorial board of international journals and conferences, including the IEEE Transactions on Control Systems Technology and the IEEE Control Systems Society Conference Editorial Board.
\end{IEEEbiography}\vfill

\end{document}